\documentclass[%
 reprint,
showpacs,
nofootinbib,
 amsmath,amssymb,
 aps,
prd,
notitlepage,
onecolumn,
superscriptaddress]{revtex4-2}

\usepackage{graphicx}
\usepackage{dcolumn}
\usepackage{bm}

\usepackage{amstext,amsmath,amssymb,amsfonts,bbm,euscript,color,dsfont}
\usepackage[latin1]{inputenc}
\usepackage{epsfig}
\usepackage{hyperref}
\usepackage{amsthm}
\usepackage{color}
\usepackage{xcolor}
\usepackage{multirow}

\usepackage{verbatim} 
\usepackage{graphicx}
\usepackage{latexsym}
\usepackage{array}
\usepackage{cleveref}
\usepackage{subcaption}
\usepackage{ulem}

\newcommand{\bea}{\begin{eqnarray}}	
\newcommand{\eea}{\end{eqnarray}}
\newcommand{\be}{\begin{equation}}	
\newcommand{\ee}{\end{equation}}
\newcommand{\beq}{\begin{equation}}	
\newcommand{\eeq}{\end{equation}}

\newcommand{\Z}{{\mathbb Z}}
\newcommand{\C}{{\mathbb C}}

\newcommand{\vev}[1]{\left\langle{#1}\right\rangle}

\newcommand{\dd}{{\textrm{d}}}

\def\R{\relax\ifmmode {\mathbb R}  \else${\mathbb R}$\fi}
\def\C{\relax\ifmmode {\mathbb C}  \else${\mathbb C}$\fi}
\def\Z{\relax\ifmmode {\mathbb Z}  \else${\mathbb Z}$\fi}
\def\N{\relax\ifmmode {\mathbb N}  \else${\mathbb N}$\fi}
\def\I{\relax\ifmmode {\mathbb I}  \else${\mathbb I}$\fi}

\begin{document}


\title{The ghost-gluon vertex in the presence of the Gribov horizon: general kinematics}  


\author{N.~Barrios}\email{nbarrios@fing.edu.uy}

\affiliation{
Instituto de F\'\i sica, Facultad de Ingenier\'\i a, Universidad de la Rep\'ublica,
J.H. y Reissig 565, 11000 Montevideo, Uruguay}\affiliation{UERJ $-$ Universidade do Estado do Rio de Janeiro,\\
Departamento de F\'isica Te\'orica, Rua S\~ao Francisco Xavier 524,\\
20550-900, Maracan\~a, Rio de Janeiro, Brasil}

\author{M.~S.~Guimaraes}\email{msguimaraes@uerj.br}

\affiliation{UERJ $-$ Universidade do Estado do Rio de Janeiro,\\
Departamento de F\'isica Te\'orica, Rua S\~ao Francisco Xavier 524,\\
20550-900, Maracan\~a, Rio de Janeiro, Brasil}

\author{B.~W.~Mintz}\email{brunomintz@gmail.com} 

\affiliation{UERJ $-$ Universidade do Estado do Rio de Janeiro,\\
Departamento de F\'isica Te\'orica, Rua S\~ao Francisco Xavier 524,\\
20550-900, Maracan\~a, Rio de Janeiro, Brasil}

\author{L.~F.~Palhares}\email{leticia.palhares@uerj.br}

\affiliation{UERJ $-$ Universidade do Estado do Rio de Janeiro,\\
Departamento de F\'isica Te\'orica, Rua S\~ao Francisco Xavier 524,\\
20550-900, Maracan\~a, Rio de Janeiro, Brasil}

\author{M.~Pel\'aez}\email{mpelaez@fing.edu.uy}

\affiliation{
Instituto de F\'\i sica, Facultad de Ingenier\'\i a, Universidad de la Rep\'ublica,
J.H. y Reissig 565, 11000 Montevideo, Uruguay}

\date{\today}

\begin{abstract}

Correlation functions are important probes for the behavior of quantum field theories. Already at tree-level, the Refined Gribov Zwanziger (RGZ) effective action for Yang-Mills theories provides a good approximation for the gluon propagator, as compared to that calculated by nonperturbative methods such as Lattice Field Theory and Dyson-Schwinger Equations. However, the study of higher correlation functions of the RGZ theory is still at its beginning. In this work we evaluate the ghost-gluon vertex function in Landau gauge at one-loop level, in $d=4$ space-time dimensions for the gauge groups SU(2) and SU(3). More precisely, we extend the analysis conducted in \cite{Mintz:2017qri} for the soft-gluon limit to an arbitrary kinematic configuration. We introduce renormalization group effects by means of a toy model for the running coupling and investigate the impact of such a model in the ultraviolet tails of our results. We find that RGZ results match fairly closely those from lattice simulations, Schwinger-Dyson equations and the Curci-Ferrari model for three different kinematic configurations. This is compatible with RGZ being a feasible theory for the strong interaction in the infrared regime.

\end{abstract}

\maketitle


\section{\label{sec:Intro}Introduction}

Since the proposal of Quantum Chromodynamics (QCD) as the theory of Strong Interactions, a long path was constructed to connect the fundamental degrees of freedom -- quarks and gluons -- to the observed physical states and processes. At high energies, asymptotic freedom allows for a perturbative approach that, supplemented by essential nonperturbative information in Particle Distribution Functions and fragmentation phenomena, agrees with a plethora of experimental output from high-energy particle colliders. The infrared (IR) regime however is much less amenable. Monte Carlo simulations that solve the Euclidean version of  QCD on a discretized space-time 
lattice  have by now and with great effort established that this non-Abelian theory can quantitatively describe several hadronic observables \cite{Fodor:2012gf,BMW:2008jgk,Borsanyi:2020mff,BMW:2014pzb}. Nevertheless, the mechanism of color confinement is still an open question, calling for the development of continuum approaches and (semi)analytical descriptions of the infrared behavior of Strong Interactions. Among the well-developed continuum methods that try to tackle this non-perturbative regime, Schwinger-Dyson equations \cite{Bashir:2012fs,Aguilar:2015bud,Aguilar:2022thg,Aguilar:2021uwa,Aguilar:2021okw,Aguilar:2021lke,Huber:2021yfy,Aguilar:2021okw,Fischer:2020xnb,Huber:2020keu,Aguilar:2018csq} stand out in different hadronic applications, but also 
the Functional 
Renormalization Group 
\cite{Berges:2000ew,Pawlowski:2005xe,Corell:2018yil,Cyrol:2016tym,Fischer:2008uz,Dupuis:2020fhh}
and effective models 
\cite{Nambu-JonaLasinio1961,Klevansky:1992qe,GellMann:1960np,Fukushima-PNJL-2003,Schaefer:2007pw} have been employed with partial success. Other approaches, such as the Curci-Ferrari (CF) model in Landau gauge \cite{Tissier:2010ts,Tissier:2011ey,Pelaez:2013cpa,Pelaez:2014mxa,Pelaez:2015tba,Gracey:2019xom,Pelaez:2021tpq,Figueroa:2021sjm,Barrios:2020ubx,Barrios:2021cks,Barrios:2022hzr} and the screened massive expansion \cite{Siringo:2022hmf,Siringo:2022dzm,Comitini:2021kxj,Comitini:2020ozt,Siringo:2018uho} have successfully described some aspects of the IR of Yang-Mills theory by employing perturbation theory. 

Here we adopt another continuum approach to the nonperturbative regime of Yang-Mills theories: the Refined Gribov-Zwanziger (RGZ) theory \cite{Dudal:2008sp,Dudal:2010tf}.
This framework, as the other continuum methods, adopts a gauge-fixed setup and is formulated from first-principles as a gauge path integral modified in the infrared by the existence of Gribov gauge copies. This idea follows the seminal work by Gribov himself \cite{Gribov:1977wm}  and the development of local actions attained by Zwanziger \cite{Zwanziger:1989mf} and complemented by the emergence of dimension two condensates \cite{Dudal:2008sp,Dudal:2010tf,Dudal:2019ing}.
For comprehensive reviews, the reader is referred to \cite{Vandersickel:2012tz,Vandersickel:2011zc,Sobreiro:2005ec} and references therein.

The presence of this nonperturbative background stemming from the Gribov horizon and the condensates seems to carry plenty of information from the interacting theory, so that the remaining interaction corrections might be supposed to be small, \textit{i.e.} perturbative.
Even at tree-level (but also at 1-loop level, as discussed in \cite{deBrito:2024ffa}), the RGZ gluon propagator is compatible with the deep IR behavior observed on Landau-gauge Lattice QCD data \cite{Dudal:2008sp}, while reducing to pure Yang-Mills at large energies \cite{Capri:2015mna} in  a fully self-consistent  formulation that would include solving the Gribov gap equation and the effective potential for the different condensates. Gauge-parameter independence of gauge-invariant quantities in this setup is under control within linear covariant gauges through a nilpotent Becchi-Rouet-Stora--Tyutin symmetry, and the predictions for the gluon propagator in general covariant gauges are also compatible with available lattice results \cite{Capri:2015ixa,Capri:2015nzw,Capri:2016aqq,Capri:2016gut}.

In this paper, the aim is to take a step forward in the direction of establishing predictions of the RGZ theory by including radiative corrections and confronting them with lattice data and other nonperturbative approaches. This is crucial for further understanding whether this theory can be a consistent model of infrared Yang-Mills and eventually QCD. One-loop corrections to the RGZ gluon propagator have been recently reported in \cite{deBrito:2024ffa}, whilw a corrected propagator for a scalar field coupled to the RGZ action has been studied in \cite{deBrito:2023qfs}. Previous results on the 
vertices of the (R)GZ action in the Landau gauge may be found in Ref. \cite{Gracey:2012wf}. Here, in particular, we compute the general kinematics of the ghost-gluon vertex at one loop order within the RGZ framework, extending previous work on the soft gluon limit of the same correlation function \cite{Mintz:2017qri}. The present study enables the investigation of the role played by auxiliary fields in quantitative results in RGZ. In this specific
correlation function, we show that the contribution of auxiliary fields, in the form of new vertices and mixed propagators, represents a smaller effect -- less than $10\%$ percent -- as compared to that of the presence of nonzero pole masses for the gluon.
Moreover, we include running effects via an infrared model, with a freezing coupling constant. Even though this is not self-consistently obtained within RGZ, this may be a reasonable approximation, because of the presence of the massive parameters from the horizon and the condensates. Overall we show that the RGZ results are fully compatible with available lattice data for SU(2) and SU(3), as well as other continuum approaches, such as the infrared-safe Curci-Ferrari model in Landau gauge \cite{Pelaez:2013cpa,Barrios:2020ubx} and the Dyson-Schwinger equations \cite{Aguilar:2018csq}.

This paper is organized as follows.
In Sec. \ref{sec:RGZ-action}, the formalism of the refined Gribov-Zwanziger framework is presented. In Sec. \ref{sec:3-point}, the ghost-gluon vertex is defined and its general structure in the RGZ theory is discussed. Our computation of generic momentum configurations of the one-loop ghost-gluon vertex is presented in Sec. \ref{sec:discussion}, including the IR running coupling model adopted.
Sec. \ref{sec:results} collects our results along with the corresponding analysis and comparisons with alternative approaches. In Sec. \ref{sec:g_toy_UV} we discuss the potential influence of the IR running coupling model on the ultraviolet (UV) tails of our results. We have included perspectives and final remarks in Section \ref{sec:finalremarks}.

\section{\label{sec:RGZ-action}The Refined Gribov-Zwanziger action in Landau gauge}

The Gribov-Zwanziger theory is a framework for making sense of the gauge fixed Yang-Mills theory in the non-perturbative regime. It grew from Gribov's observation that gauge fixing in the Landau Gauge fails in the strongly coupled regime \cite{Gribov:1977wm, Zwanziger:1989mf} (see also \cite{Vandersickel:2011zc, Vandersickel:2012tz, Sobreiro:2005ec} for reviews), in the sense that gauge copies still exist after the fixing. These are called Gribov copies and they appear in fact in any covariant gauge. The solution proposed by Gribov was to restrict the integration region of the gauge field path integral to the neighborhood containing the perturbative vacuum and bounded by the field configurations associated with the first Gribov copies, called the Gribov Horizon. In the Landau gauge $\partial_{\mu} A^a_{\mu} = 0$ the Gribov horizon is defined by the set of fields $A^a_{\mu}$ for which the equation ${\cal M}^{ab}(A)\alpha^b = 0$ has a solution, where
\begin{equation}
{\cal M}^{ab}(A)(\bullet)=-\delta^{ab}\partial^2(\bullet) +gf^{abc}\partial_{\mu}(A^{c}_{\mu} \bullet).
\label{npbrst6}
\end{equation}
stands for the Faddeev-Popov operator in the Landau gauge. 

Zwanziger \cite{ Zwanziger:1989mf} was able to recast Gribov's construction in the form of an effective theory whose restriction to the Gribov horizon is implemented at the level of the action. The resulting action is known as the nonlocal Gribov-Zwanziger action and defined in the Landau gauge in dimension $D$ by
\begin{equation}
\label{nlGZ}
S_{nlGZ} = \int d^D x\; \left( \frac 14 F^a_{\mu\nu}  F^a_{\mu\nu}  + i b^a\partial_{\mu}A^a_{\mu} - \bar{c}^a {\cal M}(A)^{ab} c^b  \right)  + \gamma^D H(A) - \gamma^D  VD\left(N^2 - 1 \right) 
\end{equation}
where $\gamma$ has dimension of mass, $V$ is the space-time volume (formally infinite) and $H(A)$ is known as the horizon function and given by
 \begin{eqnarray}
  H(A) = g^2  \int d^D p  \int d^D q\left[   f^{abd}\tilde{A}^{d}_{\mu}(-p) \left(   {\cal M}^{-1} \right)^{bc}_{pq}  f^{cae}\tilde{A}^{e}_{\mu}(q)  \right]
  \end{eqnarray} 
The mass scale $\gamma$ is not arbitrary, but fixed by a self consistent equation known as the gap equation, which can be written as a condition on a vacuum condensate. Defining 
\begin{equation}
\label{nlGZpart}
Z(\tilde\lambda) = \int {\cal D} A {\cal D} b {\cal D} \bar{c} {\cal D} c \; e^{-\int d^D x\; \left( \frac 14 F^a_{\mu\nu}  F^a_{\mu\nu}  + i b^a\partial_{\mu}A^a_{\mu} - \bar{c}^a {\cal M}(A)^{ab} c^b  \right)  + \tilde\lambda H(A) - \tilde\lambda  VD\left(N^2 - 1 \right)} 
\end{equation}
The gap equation reads
\begin{equation}
\label{gap}
\frac{\partial Z(\tilde\lambda)}{\partial \tilde\lambda}\Bigr\rvert_{\tilde\lambda = \gamma^D} = 0 \Rightarrow  \langle H(A) \rangle  = VD\left(N^2 - 1 \right).
\end{equation}

The theory as it is is highly non-local due to explicit expression of $H(A)$. In order to have a workable quantum field theory we need to put the action in a local form. This entails the introduction of new, auxiliary fields. We note that
\begin{equation}
\label{horizon-loc}
\int  {\cal D}\varphi  {\cal D}\bar{\varphi}  {\cal D}\omega  {\cal D}\bar{\omega} \;  e^{-\int\,d^4x \left(\bar\varphi^{ac}_\mu{\cal M}(A)^{ab}\varphi^{bc}_\mu - \bar\omega^{ac}_\mu{\cal M}(A)^{ab}\omega^{bc}_\mu
+ ig\gamma^{\frac D2}\,f^{abc}A^a_\mu(\varphi_\mu^{bc} + \bar\varphi^{bc}_\mu)\right)} \sim e^{-\gamma^D\;  H(A) }
 \end{equation}
where the symbol $\sim$ means ``up to a  prefactor'' and  $(\varphi, \bar{\varphi} )$ are complex bosonic fields and  $(\omega, \bar{\omega} )$ are fermionic fields. Therefore we obtain the local formulation as 
\begin{equation}
\label{GZlocalpartition}
Z_{GZ}  = \int  {\cal D}A {\cal D}\bar{c}  {\cal D}c  {\cal D}b  {\cal D}\bar{\varphi} {\cal D}\varphi  {\cal D}\bar{\omega}  {\cal D}\omega\;  e^{-S_{GZ} }
 \end{equation}
where
\begin{align}
\label{GZaction}
S_{GZ} = &\int d^D x\; \left( \frac 14 F^a_{\mu\nu}  F^a_{\mu\nu}  + i b^a\partial_{\mu}A^a_{\mu} - \bar{c}^a {\cal M}(A)^{ab} c^b   \right)  \nonumber\\
&+ \int\,d^Dx \left(\bar\varphi^{ac}_\mu{\cal M}(A)^{ab}\varphi^{bc}_\mu - \bar\omega^{ac}_\mu{\cal M}(A)^{ab}\omega^{bc}_\mu \right)\nonumber\\
&+ \int d^Dx   \left( ig\gamma^{\frac D2}\,f^{abc}A^a_\mu(\varphi_\mu^{bc} + \bar\varphi^{bc}_\mu) - \gamma^D  D\left(N^2 - 1 \right)  \right)
\end{align}
and $\gamma$ is such that it satisfies the gap equation \eqref{gap}, which now reads   
\begin{align}
\label{gapequation2}
g   \langle \,f^{abc}A^a_\mu(\varphi_\mu^{bc} + \bar\varphi^{bc}_\mu) \rangle =   \gamma^{\frac D2}  D\left(N^2 - 1 \right)  
\end{align}
where translation invariance of the vacuum was used to factor out the volume $\int d^D x \langle \,f^{abc}A^a_\mu(\varphi_\mu^{bc} + \bar\varphi^{bc}_\mu) \rangle = V \langle \,f^{abc}A^a_\mu(\varphi_\mu^{bc} + \bar\varphi^{bc}_\mu) \rangle $. 

The vacuum defined by the gap equation is unstable and favors the formation of new condensates \cite{Dudal:2008sp} that are related to mass scales of the gauge fields $\langle A^2 \rangle \sim m^2$ and auxiliary fields $\langle \bar{\phi} \phi \rangle \sim \mu^2$. The incorporation of these condensates in the effective action formulation leads to a modified theory known as the Refined Gribov-Zwanziger theory, defined by  
\begin{align}
\label{RGZaction}
S_{RGZ} = &\int d^D x\; \left( \frac 14 F^a_{\mu\nu}  F^a_{\mu\nu}  + i b^a\partial_{\mu}A^a_{\mu} - \bar{c}^a {\cal M}(A)^{ab} c^b   \right)  \nonumber\\
&+ \int\,d^Dx \left(\bar\varphi^{ac}_\mu{\cal M}(A)^{ab}\varphi^{bc}_\mu - \bar\omega^{ac}_\mu{\cal M}(A)^{ab}\omega^{bc}_\mu \right)\nonumber\\
&+ \int d^Dx   \left( ig\gamma^{\frac D2}\,f^{abc}A^a_\mu(\varphi_\mu^{bc} + \bar\varphi^{bc}_\mu) - \gamma^D  D\left(N^2 - 1 \right)  \right)\nonumber\\
&+ \int\dd^Dx~\frac{m^2}{2}A_\mu^aA_\mu^a + \int\dd^Dx \;M^2\left( \bar{\varphi}^{bc}_{\mu}\varphi^{bc}_{\mu}- 
 \bar{\omega}^{bc}_{\mu} \omega^{bc}_{\mu}\right).
\end{align}

It is worth pointing out the remarkable fact discovered in \cite{Capri:2016aqq} that it is possible to recast this theory making it suitable for any linearly covariant gauge, so that the resulting RGZ theory can be made BRST invariant. This is done by replacing the gauge field $A$ by a gauge invariant composite field $A^h$ \cite{Lavelle:1995ty,Capri:2016aqq} 
\begin{equation}\label{eq:def-Ah-local}
A^{h}_{\mu}=h^{\dagger}A_{\mu}h+\frac{i}{g}h^{\dagger}\partial_{\mu}h\,,
\end{equation}
with
\begin{equation}\label{eq:def-h}
h=\mathrm{e}^{ig\xi^a T^a}\equiv \mathrm{e}^{ig\xi},
\end{equation}
where $\xi^a$ is the Stueckelberg field discussed in 
\cite{Lavelle:1995ty,Fiorentini:2016rwx,Capri:2015nzw,
Capri:2015ixa,Capri:2016aqq,Capri:2016gut}. One also 
needs to impose a transversality constraint on $A^h$,
\begin{eqnarray}\label{eq:dAh}
\partial_{\mu}A^{h,a}_{\mu}=0\,,
\end{eqnarray} 
so that the $h$ field is not really independent, but is an auxiliary field. This leads to a consistent renormalizable and BRST invariant formulation. This enlarged formulation reduces to the Landau gauge if $\xi = 0$. In this work we will only consider the Landau gauge.

\section{The three-point ghost-gluon correlation function}
\label{sec:3-point} 

Let us now start discussing the ghost-gluon correlation function and establish some notation. We first make some general remarks about the connected gluon-antighost-ghost three-point function in the RGZ theory and how the mixed propagators affect its building blocks. Next, for completeness, we quote the results obtained in \cite{Mintz:2017qri} for the soft-gluon limit. Our original results for the gluon-antighost-ghost correlation function for arbitrary momenta are left for the next section.

\subsection{General structure of the $A\bar cc$ vertex}

The relevant Feynman rules for the calculation 
of the ghost-gluon vertex at one-loop order can be derived from the action  (\ref{RGZaction}) and are listed in Appendix \ref{sec:appendix-feynman-rules}. 

With these rules at hand, one can calculate the connected correlation function
\begin{eqnarray}\label{eq:cbcA-def}
 \vev{A_\mu^a(k)\,\bar c^b(p)\,c^c(q)}_{q=-p-k} = \left.\frac{\delta^3Z_c[J_A,J_{\bar c},J_c]}{\delta (J_{\bar A})_\mu^a(k)\delta J_{\bar c}^b(p)\delta J_{c}^c(q)}\right|_{q=-p-k;\;\;\;J_i=0}
\end{eqnarray}
at one-loop order, where $Z_c$ is the generator of connected 
correlation functions and $J_i$ 
($i=\bar c,c,A$) are external sources linearly coupled to the fields $i$. As usual, the sources are taken to zero at the end of the calculation.


\begin{figure}[b]
\includegraphics[height=6cm]{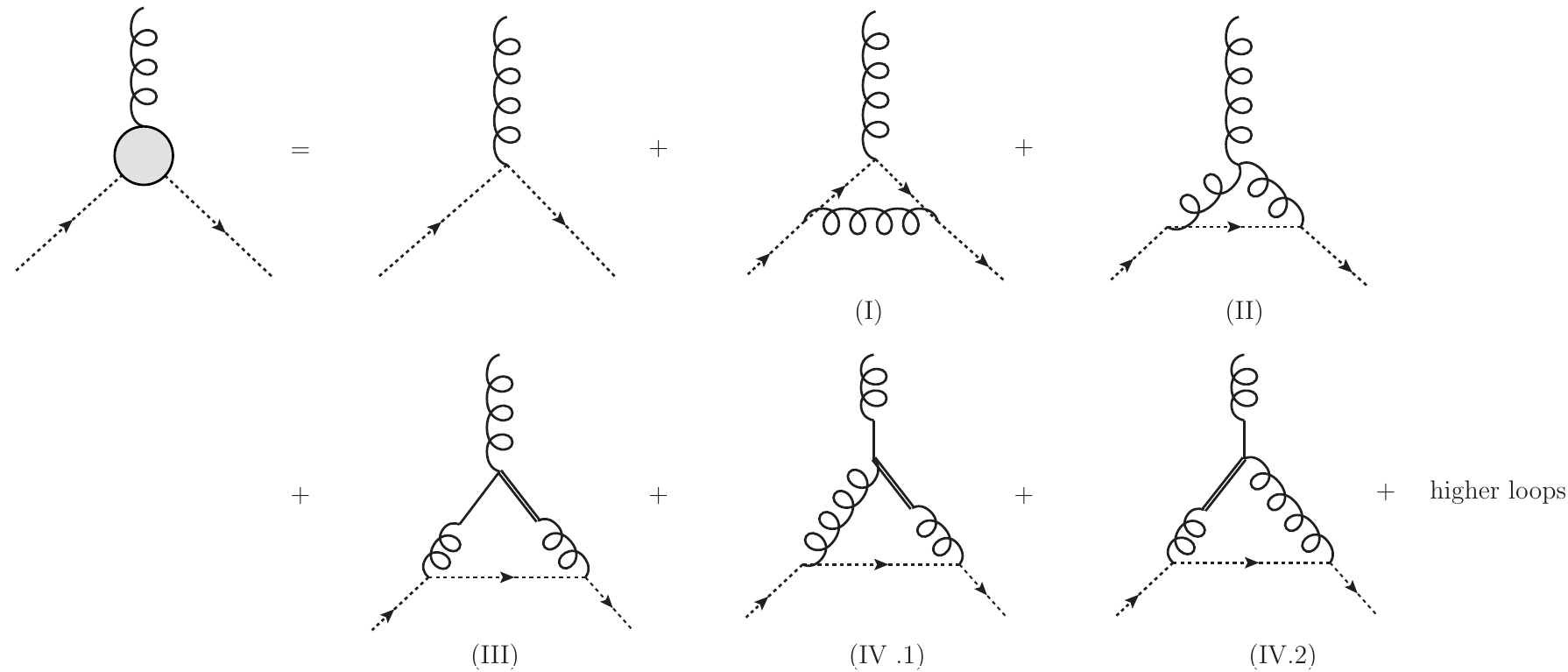}
\caption{Feynman diagram expansion up to one-loop order for the ghost-gluon vertex in the Refined Gribov-Zwanziger theory. Dashed lines represent ghosts and antighosts, while the curly lines stand for gluons. Full simple and double lines, that only appear in mixed propagators, correspond to the auxiliary fields $\varphi$ and $\bar\varphi$, respectively. Of course, vertices are evaluated at tree-level order by employing the expressions provided in \cref{sec:appendix-feynman-rules}. Note that as for the diagrams (IV.1) and (IV.2) the external leg involving the gluon corresponds to a mixed propagator of the form $\vev{A\varphi}$. The roman numbers identifying the one-loop diagrams will be used as reference in the results section and in the appendices. The diagrams were generated by means of the \textsc{Jaxodraw} interface \cite{Binosi:2008ig}.}\label{fig:feyndiags}
\end{figure}

Before proceeding, let us remark that since the RGZ action 
contains bilinear couplings between fields, the theory contains 
mixed propagators, such as $\vev{A\varphi}$ and 
$\vev{A\bar\varphi}$. Therefore, the relation between 
connected and 1PI functions has to take such mixed propagators 
into account. This is made explicit in the Feynman diagrams of 
Fig. \ref{fig:feyndiags}. Such mixed propagators and vertices 
involving Zwanziger's auxiliary fields $\varphi$ and 
$\bar\varphi$ as well as their  fermionic counterparts 
$\omega$ and $\bar\omega$, arise as a consequence of the local 
formulation of the Gribov horizon \footnote{Another possible formulation of the theory would include 
nonlocal, momentum-dependent vertices instead of the auxiliary 
fields. However, for the sake of using standard Quantum Field Theory 
techniques, we employ the  local version of the theory.}. We give 
further details for the interested reader in Appendix 
\ref{sec:appendix-mixed-propagators}.

Since the tree-level mixed propagator is such that
\begin{eqnarray}\label{eq:Aphi-propagator}
\langle A_{\mu}^a(p)\varphi_{\nu}^{bc}(-p)\rangle
&=&\frac{-ig\gamma^2f^{abc}}{p^4+p^2(m^2+M^2)+m^2M^2+2Ng^2\gamma^4}P^{\perp}_{\mu\nu}(p),
\end{eqnarray}
the one-loop connected function (\ref{eq:cbcA-def}) can then be decomposed as\footnote{This expression has already been derived in \cite{Mintz:2017qri}. However, the result showed here differs from it by a factor of (-2i) in the coefficient of $\Gamma_{\bar c^b c^c \varphi_{\mu}^{de}}$ as a consequence of a different choice for the Feynman rules. Of course, this is just a matter of convention and does not impact on the evaluation of the vertex function.}
\begin{align}
 \vev{A_\nu^a(k)\,\bar c^b(p)\,c^c(q)} &=  G(p)G(q)D_{AA}(k)P^{\perp}_{\mu\nu}(k)\left\{\Gamma_{A^a_\mu\bar c^b c^c}(k,p,r)
 -
 \frac{2i g\gamma^2f^{ade}}{k^2+M^2}\Gamma_{\bar c^b c^c \varphi^{de}_\mu }(k,p,r)\right\}_{q=-p-k},
 \label{eq:Acc-local-shorthand-notation}
\end{align}
where
\begin{eqnarray}
 P^{\perp}_{\mu\nu}(p) = \delta_{\mu\nu} - \frac{p_\mu p_\nu}{p^2}
\end{eqnarray}
is the transverse projector, and the relevant 1PI functions are
\begin{eqnarray}
    \Gamma_{A^a_\mu\bar c^b c^c}(k,p,r):=\frac{\delta^3\Gamma}{\delta A_\mu^a(-k)\delta \bar c^b(-p)\delta  c^c(-q)}
\end{eqnarray}
and
\begin{eqnarray}
    \Gamma_{\bar c^b c^c \varphi^{de}_\mu }(k,p,r):=\frac{\delta^3\Gamma}{\delta \bar c^b(-p)\delta  c^c(-q)\delta \varphi_\mu^{de}(-k)}.
\end{eqnarray}

Note the presence of the $\Gamma_{\bar c c \varphi}$ and $\Gamma_{\bar c c \bar\varphi}$ vertex functions in the RGZ theory, which are zero at tree-level order. In the case of $\Gamma_{\bar c c \varphi}$ its first nontrivial terms are given by diagrams (IV.1) and (IV.2) of \cref{fig:feyndiags}, where the external leg involving the gluon field corresponds to a propagator of the type $\vev{A \varphi}$ connected to a $\bar{\varphi} \varphi A$ tree level vertex. In contrast, the vertex functions $\Gamma_{\bar c c \omega}$ and $\Gamma_{\bar c c \bar\omega}$ do not contribute to the ghost-gluon vertex due to the absence of mixed propagators of the type $\vev{A \omega}$ within the RGZ theory. Notice also that, since $\Gamma_{\bar c c \varphi}=\Gamma_{\bar c c \bar\varphi}$, in \cref{fig:feyndiags} we have included the diagrams associated with the former quantity only. The contribution of $\Gamma_{\bar c c \bar\varphi}$ to $\vev{A_\nu^a(k)\,\bar c^b(p)\,c^c(q)}$ is introduced by multiplying the contribution of $\Gamma_{\bar c c \varphi}$ by a factor of two, as can be observed in \cref{eq:Acc-local-shorthand-notation}.

The tensorial structure of the ghost-gluon 1PI vertex function is given by
\begin{eqnarray}
  \Gamma_{A^a_\mu\bar c^b c^c}(k,p,r)=-igf^{abc}\large[p_\mu B_1(k,p) + k_\mu B_2(k,p)\large],
  \label{eq:Acbarc-tensor-decomp}
\end{eqnarray}
where we adopt the same notation as in \cite{Aguilar:2018csq} and consider all momenta as incoming. At tree-level, the scalar structure functions are $B_1(k,p)=1$ and $B_2(k,p)=0$.

On the other hand, the contraction of the $\Gamma_{\varphi\bar c c}$ function with the antisymmetric color structure constant tensor can be decomposed as
\begin{eqnarray}
    f^{ade}\Gamma_{\bar c^b c^c \varphi^{de}_\mu }(k,p,r) = gf^{abc}\left(p_\mu C_1(k,p) + k_\mu C_2(k,p)\right)+gd^{abc}\left(p_\mu C'_1(k,p) + k_\mu C'_2(k,p)\right)
\end{eqnarray}
Since this combination must be antisymmetric in the color indices $b$ and $c$ (due to the fact that the ghost fields are grassmannian), the coefficient of the totally symmetric tensor $d^{abc}$ must vanish identically, \textit{i.e.}, $C'_1=C'_2=0$. As a result, we may write the one-loop connected three-point function as
\begin{eqnarray}
 \vev{A_\mu^a(k)\,\bar c^b(p)\,c^c(q)} &=&  
 -igp_\nu\,f^{abc}G(p)G(p+k)D_{AA}(k)P^{\perp}_{\mu\nu}(k)
 \left\{ B_1(k,p) +
 \frac{2g\gamma^2}{k^2+M^2} C_1(k,p)
 \right\},
\end{eqnarray}
where the longitudinal scalar functions $B_2$ and $C_2$ are no longer present (in the Landau gauge) due to the transversality of the gluon propagator.

In order to make contact with results from Monte Carlo lattice simulations, let us consider the scalar quantity \cite{Cucchieri:2006tf}
\begin{align}
    G^{CCA}(p,k)&\equiv\frac{\Gamma_{(A\bar c c, tree)}^{abc} \,_\mu(k,p,-k-p) \vev{A_\mu^a(k)\,\bar c^b(p)\,c^c(-k-p)}}{\Gamma_{(A\bar c c, tree)}^{abc} \,_\mu(k,p,-p-k)\Gamma_{(A\bar c c, tree)}^{abc} \,_\nu(k,p,-p-k)P_{\mu\nu}(k)G(p)G(p+k)D_{AA}(k)}\nonumber\\
    &= B_1(k,p) + \frac{2 g\gamma^2}{k^2+M^2}C_1(k,p),
    \label{eq:GCCA_B1_C1}
\end{align}
which from now on we denote as the ghost-gluon vertex dressing function.

There are clearly some differences between perturbative YM and RGZ calculations 
of the vertex function. The first of them is the modification of the gluon propagator 
brought about by the restriction to the Gribov horizon, which can be 
understood as the appearance of a pair of generally complex conjugate poles.  A second difference is the presence of the tree-level 
$A\bar\varphi\varphi$ vertex, which couples the gluon to the auxiliary Zwanziger fields. 
This allows not only diagrams with auxiliary fields running in the internal loops, but also 
in the external legs, as long as the external propagator is a mixed one like, for example,  $\vev{A\varphi}$. This possibility is realized in (\ref{eq:Acc-local-shorthand-notation}), 
giving rise to the contributions $\Gamma_{\varphi\bar c c}$ 
and $\Gamma_{\bar\varphi\bar c c}$, not present in perturbative YM theory. Furthermore, note 
that these mixed contributions only appear from one-loop order onwards, as such vertices
are absent from the classical action (\ref{RGZaction}). Being finite, they do not spoil the stability of the action, in agreement with the Quantum Action Principle \cite{Piguet:1995er}. Finally, we note that the presence of such vertices involving the auxiliary Zwanziger fields can be thought as effective momentum-dependent gluonic interaction terms. This claim can be explicitly shown to be true, starting from the tree level action, by considering the nonlocal version of the RGZ theory (whose action can be obtained from (\ref{RGZaction}) by integrating out the Zwanziger auxiliary fields)  and expanding the inverse of the Faddeev-Popov operator in powers of the coupling. In this regard, we note that it is reasonable to consider the RGZ effective action not simply as an effective model with propagators given by a ratio of polynomials, but rather a renormalizable effective theory with nontrivial gluonic propagators and self-interactions, which are related to each other.

As a final remark, let us note that all diagrams contributing to the ghost-gluon vertex are finite in the RGZ theory, just as in the perturbative Yang-Mills theory. In fact, the mixed propagators present in diagrams (III) and (IV) make the diagrams even more ultraviolet convergent than the usual YM diagrams (I) and (II). This can be easily seen from the form of the $\vev{A\varphi}$ propagator (\ref{eq:Aphi-propagator}).

\subsection{The soft gluon limit}\label{sec:soft-gluon}

The calculation of the ghost-gluon vertex for any momentum requires the calculation of the diagrams shown in \cref{fig:feyndiags}. Since this calculation is rather long to be performed manually, in a previous work some of us considered the soft-gluon limit of $\Gamma_{A\bar c c}$, \textit{i.e.}, the limit in which the gluon momentum $k\rightarrow0$ \cite{Mintz:2017qri}. 

In the soft gluon limit, each diagram takes a simplified form. Moreover, diagrams (IV.1) and (IV.2) vanish. In reality, as for the soft gluon limit, these contributions vanish at all orders of perturbation theory. In the case of $\Gamma_{\bar c^b c^c \varphi_\mu^{de}}$, this is due to the derivative nature of the $\varphi \bar\varphi A$ vertex and the transversality of the gluon propagator, as well as all propagators involving the $\varphi$ field (see \cref{sec:appendix-feynman-rules}). This property could also be inferred by looking at the function $\Gamma_{\bar c^b c^c \bar\varphi_\mu^{de}}$ which is identical to $\Gamma_{\bar c^b c^c \varphi_\mu^{de}}$. As $\Gamma_{\bar c^b c^c \bar\varphi_\mu^{de}}$ is always proportional to the momentum of the external gluon, it vanishes as this momentum goes to zero\footnote{In both cases of $\Gamma_{\bar c^b c^c \varphi_\mu^{de}}$ and $\Gamma_{\bar c^b c^c \bar\varphi_\mu^{de}}$ we are considering loop corrections to be regular as $k\to 0$. This is indeed the case of the RGZ framework, owing to the fact that the parameters $m$, $M$ and $\gamma$ act as natural IR regulators of the theory.}.

In this limit, the one-loop correction for the vertex, which has been explicitly calculated in \cite{Mintz:2017qri}, reads
\begin{eqnarray}\label{eq:resultGamma}
 [\Gamma_{A\bar c c}^{(1)}(0,p,-p)]^{abc}_{\mu} 
&=&ig^3\frac{Nf^{abc}}{2}\bigg\{R_+ J_\mu(a_+;p) + R_- J_\mu(a_-;p)
+2R_+^2K_\mu(a_+,a_+;p) + 2R_-^2K_\mu(a_-,a_-;p) + \nonumber\\
&&\left.+4R_+R_- K_\mu(a_+,a_-;p) +
\frac{N}{2}\left(\frac{g\gamma^2}{a_+^2-a_-^2}\right)^2
\left[K_\mu(a_+,a_+;p)+K_\mu(a_-,a_-;p)-\right.\right.\nonumber\\
&&-2K_\mu(a_+,a_-;p)\big]\bigg\}\,,
\end{eqnarray}
where
\begin{align}
 & a_+^2 \equiv \frac{m^2+M^2+\sqrt{(m^2-M^2)^2-4\lambda^4}}{2},\nonumber\\
 & a_-^2 \equiv \frac{m^2+M^2-\sqrt{(m^2-M^2)^2-4\lambda^4}}{2},\nonumber
 \end{align}
with $\lambda^4\equiv 2 N g^2 \gamma^4$, and the integral
\begin{equation}
 J_\mu(m_1;p)\equiv\int \frac{d^d q}{(2\pi)^d}\,
\frac{1}{q^2}\frac{1}{q^2+m_1^2}\frac{p^2q^2 - (p\cdot q)^2}{[(q-p)^2]^2}\,(q-p)_\mu,
\end{equation}
is related to diagram (I) while 
\begin{eqnarray}
 K_\mu(m_1,m_2;p)&\equiv&\int \frac{d^dq}{(2\pi)^d} \frac{1}{(q+p)^2}\frac{1}{q^2+m_1^2}\frac{1}{q^2+m_2^2}
\,\left[\frac{q^2 p^2 - (p\cdot q)^2}{q^2}\right]q_\mu\,
\end{eqnarray}
appears in diagrams (II) and (III)\footnote{Analytic expressions of $J_\mu (m;p)$ and $K_\mu(m_1,m_2;p)$ are provided in \cref{sec:appendix-soft-gluon}.}. The incoming antighost momentum is given by $p$.  Therefore, the ghost momentum is $-p$, since $k=0$ in the soft-gluon limit. The massive parameters
$-a_\pm^2$ are the,  generally complex, poles of the RGZ gluon propagator (\ref{eq:RGZgluonpropagator}) and $R_\pm$ are their corresponding residues. 
It is worth pointing out that the last terms in eq. \eqref{eq:resultGamma} come from diagram (III) in \cref{fig:feyndiags} 
which is absent in standard YM theories, being proportional to the Gribov parameter. 

As a last remark, let us note an important feature of the $\Gamma_{A\bar c c}$ vertex function in the RGZ theory: it explicitly respects the so-called Taylor kinematics, {\it i.e.}, 
\begin{eqnarray}\label{eq:Taylor-kin-antighost}
 (\Gamma_{A\,\bar c\,c})^{abc}_\mu(p,0,-p) = 0\,,
\end{eqnarray}
and the so-called non-renormalization theorem of the ghost-gluon vertex 
\begin{eqnarray}\label{eq:Taylor-kin-ghost}
 (\Gamma_{A\,\bar c\,c})^{abc}_\mu(-p,p,0) = -i gf^{abc}p_\mu,
\end{eqnarray}
which are the same in the RGZ framework as in perturbative YM theory \cite{Taylor:1971ff}. These are direct
consequences of the Ward identities of the action 
(\ref{RGZaction}).

\section{The one-loop RGZ ghost-gluon vertex for generic momentum configurations} 
\label{sec:discussion}

\subsection{Feynman diagrams}

It is clear from \cref{eq:GCCA_B1_C1} that the ghost-gluon vertex dressing function relates to two 1PI structures. Diagrams (I)-(III) from \cref{fig:feyndiags} contribute to $\Gamma_{A\bar c c}$, while the forth diagram contributes to $\Gamma_{\varphi \bar c c}$. 

Their expressions in terms of the Feynman rules from \cref{sec:appendix-feynman-rules}, following the numbering from \cref{fig:feyndiags}, are:

\begin{align}
    & D_{\text{(I)}}=-ig^3\frac{N}{2}f^{abc}p_\nu(p+k)_\rho\int\frac{d^dq}{(2\pi)^d}q_\mu P^\perp_{\nu\rho}(q+p)D_{AA}(q+p)D_{\bar cc}(q-k)D_{\bar cc}(q)\,,\nonumber\\
    & D_{\text{(II)}}=-ig^3\frac{N}{2}f^{abc}p_\eta(p+k)_\omega\int\frac{d^dq}{(2\pi)^d}\left(-2k_\rho\delta_{\mu\nu}+2k_\nu\delta_{\mu\rho}+(2q-k)_\mu\delta_{\nu\rho}\right) P^\perp_{\nu\eta}(q)P^\perp_{\rho\omega}(q-k)D_{AA}(q)D_{AA}(q-k)D_{\bar cc}(q+p)\,,\nonumber\\
    & D_{\text{(III)}}=ig^3\frac{N^2}{4}f^{abc}(g\gamma^2)^2p_\nu(p+k)_\rho\int\frac{d^dq}{(2\pi)^d}(2q-k)_\mu P^\perp_{\rho\eta}(q-k)P^\perp_{\nu\eta}(q)D_{A\varphi}(q-k)D_{A\varphi}(q)D_{\bar cc}(q+p)\,,\nonumber\\
    & f^{ade}D_{\text{(IV.1)}}=ig^3\frac{N^2}{4}f^{abc}(g\gamma^2)p_\eta k_\omega (p+k)_\sigma\int\frac{d^dq}{(2\pi)^d}P^\perp_{\sigma\mu}(q-k)P^\perp_{\eta\omega}(q)D_{A\varphi}(q-k)D_{AA}(q)D_{\bar cc}(q+p)\,,\nonumber\\
    & f^{ade}D_{\text{(IV.2)}}=-ig^3\frac{N^2}{4}f^{abc}(g\gamma^2)p_\sigma k_\omega (p+k)_\eta\int\frac{d^dq}{(2\pi)^d}P^\perp_{\sigma\mu}(q)P^\perp_{\eta\omega}(q-k)D_{A\varphi}(q)D_{AA}(q-k)D_{\bar cc}(q+p)\,,
\end{align}
where $D_{XY}(q)$ designates the propagator of momentum $q$ between the fields $X$ and $Y$.

The above Feynman integrals depend on two external momenta, \textit{p} and \textit{k}, and feature only one Lorentz index which is not contracted, $\mu$. As a result, all of them can be expressed as $F(p^2,k^2,p\cdot k,M^2,m^2,\lambda) p_\mu+H(p^2,k^2,p\cdot k,M^2,m^2,\lambda) k_\mu$, where $F$ and $H$ are scalar integrals. In the first stage of our computation we found the corresponding scalar integrals for each diagram\footnote{The contributions for the dressing function $G^{CCA}$ come from the $p_\mu$ component entirely.} and evaluated the color factors by means of the \textsc{ColorMath} package \cite{Sjodahl:2012nk}.

In a second stage, we reduced the scalar integrals to one-loop master integrals \cite{Passarino:1978jh} of the type 

\begin{align}
    &A_x\equiv\int_q G_x(q^2),\\
    &B_{xy}(p^2)\equiv\int_q G_x(q^2)G_y((p-q)^2),\\
    &C_{xyz}(p^2,k^2,p\cdot k)\equiv\int_q G_x(q^2)G_y((p-q)^2)G_z((k-q)^2),
    \label{eq:one-loop_masters}
\end{align}
where
\begin{equation}
    \int_q\equiv \int  16 \pi^2 \mu^{2\epsilon} \frac{d^d q}{(2\pi)^d},
    \label{eq:integral_measure}
\end{equation}
with $\epsilon=\frac{4-d}{2}\geq 0$ and 
\begin{equation}
    G_x(q^2)\equiv\frac{1}{q^2+x},
\end{equation}
where $x$ is a ``mass'' squared (which may be complex, in our case). In the case of the RGZ theory $x\in \{0,M^2,a_+^2,a_-^2\}$. The mass dimension of the gauge coupling, defined as $\mu^{\epsilon}$, is absorbed into the definition of the master integrals. More specifically, as the dressing function $G^{CCA}$ is proportional to $g^2$, we include the respective mass dimension, $\mu^{2\epsilon}$, in the definition given by eq. (\ref{eq:integral_measure}). 

We implemented the reduction to master integrals using an algorithm in \textsc{Mathematica} based on the \textsc{Fire} package \cite{Smirnov:2019qkx}. The reduction for each diagram is presented in a supplementary material. The analysis is performed for an arbitrary number of colors $N$ with the purpose of comparing with alternative approaches, both for the SU(2) and SU(3) gauge groups. 

Once the reduction is finished we need to compute the one-loop master integrals. For both $A$ and $B$ there are well-known analytic results (see, for instance, ref. \cite{Martin:2003qz}). In the case of the master integral $C$, when needed, we evaluated it numerically.

We note here that since all one-loop diagrams are finite no renormalization factors are needed.

\subsection{Kinematic configurations}

Concerning the particular kinematic configurations we will address in this article, it is convenient to recall that three-point correlation functions depend on two external momenta. Because of translational and rotational invariance such dependence can be described by means of three independent kinematic variables. We choose the magnitudes of the antighost and gluon momentum, $p$ and $k$, along with the scalar product $\textit{p}\cdot \textit{k}$. 

In this paper we aim at extending the analysis of the ghost-gluon vertex within the RGZ framework, initially explored in \cite{Mintz:2017qri}, beyond the soft gluon limit, to include arbitrary kinematic configurations. We focus on the specific kinematic setups for which we have access to lattice simulations, namely the symmetric and orthogonal configurations. 

The symmetric configuration is characterized by equal momentum magnitudes for the external antighost and gluon legs, $p=k$, forming an angle of $\theta=\pi/3$. On the other hand, the orthogonal configuration refers to external momenta perpendicular to each other, $\theta=\pi/2$. Within this work, our reference to the orthogonal configuration includes the additional condition $p=k$. 

\subsection{Numerical values of the RGZ parameters}

Within the RGZ framework, there are four parameters: $M$ and $m$, that are mass parameters associated with dimension two condensates, the Gribov parameter $\gamma$ and the gauge coupling $g$.  For the sake of convenience, we introduce the parameter $\lambda^4=2 N g^2 \gamma^4$, which we will employ in place of $\gamma$.

In principle, the three massive parameters could be determined self-consistently using the Gribov gap equation and minimizing the effective potential of the theory with respect to two condensates. Here, however, we shall fix these parameters using lattice input.
The parameters $M, m$ and $\lambda$ at tree-level were determined for SU(2) in \cite{Cucchieri:2011ig} and for SU(3) in \cite{Oliveira:2012eh}, by fitting the lattice data of the gluon propagator from YM theory with an analytic expression of the form:

\begin{equation}
    D(p^2)=\frac{p^2+a}{p^4+b p^2+c},
    \label{eq:D_fit}
\end{equation}
where $a$, $b$ and $c$ are constants. This analytic form coincides with the tree-level gluon propagator of the RGZ framework,
\begin{equation}
    D(p^2)=\frac{p^2+M^2}{p^4+(M^2+m^2)p^2+M^2m^2+\lambda^4},
\end{equation}
allowing for the extraction of $m$, $M$ and $\lambda$ from eq. (\ref{eq:D_fit}). The obtained values are displayed in \cref{table:parameters_m_M_lambda}.

\begin{table}[h!]
  \begin{center}
    \begin{tabular}{|c||c|c|c|}
    \hline
     Gauge group & \hspace{.2cm} $M^2$ (${\rm GeV}^2$) \hspace{.2cm} & \hspace{.2cm} $m^2$ (${\rm GeV}^2$) \hspace{.2cm} & \hspace{.2cm} $\large\lambda$ (${\rm GeV}$) \\
\hline \hline
\rule{0pt}{3ex} 
 SU(2) - Ref.\cite{Cucchieri:2011ig}  & 2.51    & -1.92 &   1.52\\
 \hline
 SU(3) - Ref.\cite{Oliveira:2012eh}&   4.47  & -3.77   & 2.04 \\
 \hline
\end{tabular}  
\end{center}
    \caption{RGZ parameters fitted from lattice results for SU(2) \cite{Cucchieri:2011ig} and SU(3) \cite{Oliveira:2012eh}.}
    \label{table:parameters_m_M_lambda}
\end{table}

\subsection{Toy model of the running coupling}

The gauge coupling $g$ is to be determined by fitting the RGZ ghost-gluon vertex to available lattice simulations. To that end, it is useful to introduce the relative error between the RGZ output and the lattice data, 
\begin{equation}
    \chi^2\equiv\frac{1}{N_{\text{latt}}^2}\sum_i^{N_\text{latt}}\Bigg(\frac{G_{\text{RGZ}}(p_i)-G_{\text{latt}}(p_i)}{G_{\text{latt}}(p_i)}\Bigg)^2,
    \label{eq:rel_error}
\end{equation}
where $N_{\text{latt}}$ is the number of lattice data points, $G_{\text{RGZ}}$ and $G_{\text{latt}}$ refer to the RGZ and the numerical results, respectively.

Firstly, we simply investigate at a qualitative level the variation of the dressing function $G^{CCA}$ as we change fixed values of $g$. As we will see, this is enough to reproduce either the IR or the UV but not both regimes at the same time. Most likely this is due to the presence of large logarithms in the UV domain, which spoils the validity of perturbation theory. This problem can be overcome by introducing the running of the various parameters of the RGZ framework and choosing a renormalization scale of the same order of the external momentum: $\mu\sim p$. To achieve this we would need to evaluate the two-point functions of the theory by introducing an infrared-safe scheme, of the type introduced in \cite{Pelaez:2013cpa} for instance. 

However, this sort of approach exceeds the scope of the present article. Instead, we keep the constant values provided in \cref{table:parameters_m_M_lambda} and make use of a toy model for the renormalization group (RG) flow of the gauge coupling, motivated by the standard Yang-Mills one loop $\beta$-function in the Modified Minimal Subtraction Scheme,
\begin{equation}
    g^2(\mu)=\frac{g_0^2}{1+\frac{11}{3}N \frac{g_0^2}{16\pi^2}\log(\frac{\mu^2+\Lambda^2}{\Lambda^2})},
\label{eq:g_running}
\end{equation}
and choose $\mu=p$. The parameter $\Lambda$ is introduced to regularize the IR\footnote{The $\Lambda$ parameter is crucial so that the toy model is Landau pole free.} by freezing the running  of the gauge coupling for momentum scales much smaller than $\Lambda$, where we expect the effects of the renormalization flow to be less relevant. The fits of $G^{CCA}$ from the RGZ framework to lattice data are carried out by selecting the parameters $g_0$ and $\Lambda$ that minimize $\chi$.

\section{Results}
\label{sec:results}

In this section we show the results for the ghost-gluon vertex dressing function, $G^{CCA}$, for both the gauge groups SU(2) and SU(3) in four spacetime dimensions. We compare our results with lattice simulations and outcomes from the CF model and DSE. Additionally, we show the contributions of each Feynman diagram to the final results.

\subsection{SU(2)}

We begin by analyzing the ghost-gluon vertex dressing function $G^{CCA}$ for the SU(2) gauge group in the symmetric and orthogonal kinematic configurations, as well as in the case of the soft gluon limit. The latter case was previously investigated in \cite{Mintz:2017qri}. However, for completeness and because the analysis we perform in this article differs slightly from the one developed in \cite{Mintz:2017qri}, we include those results here as well.

\subsubsection{Fitting the ghost-gluon vertex dressing function}

\begin{figure}[h!]
    \centering
    \begin{subfigure}{0.4\textwidth}
    \includegraphics[width=\textwidth]{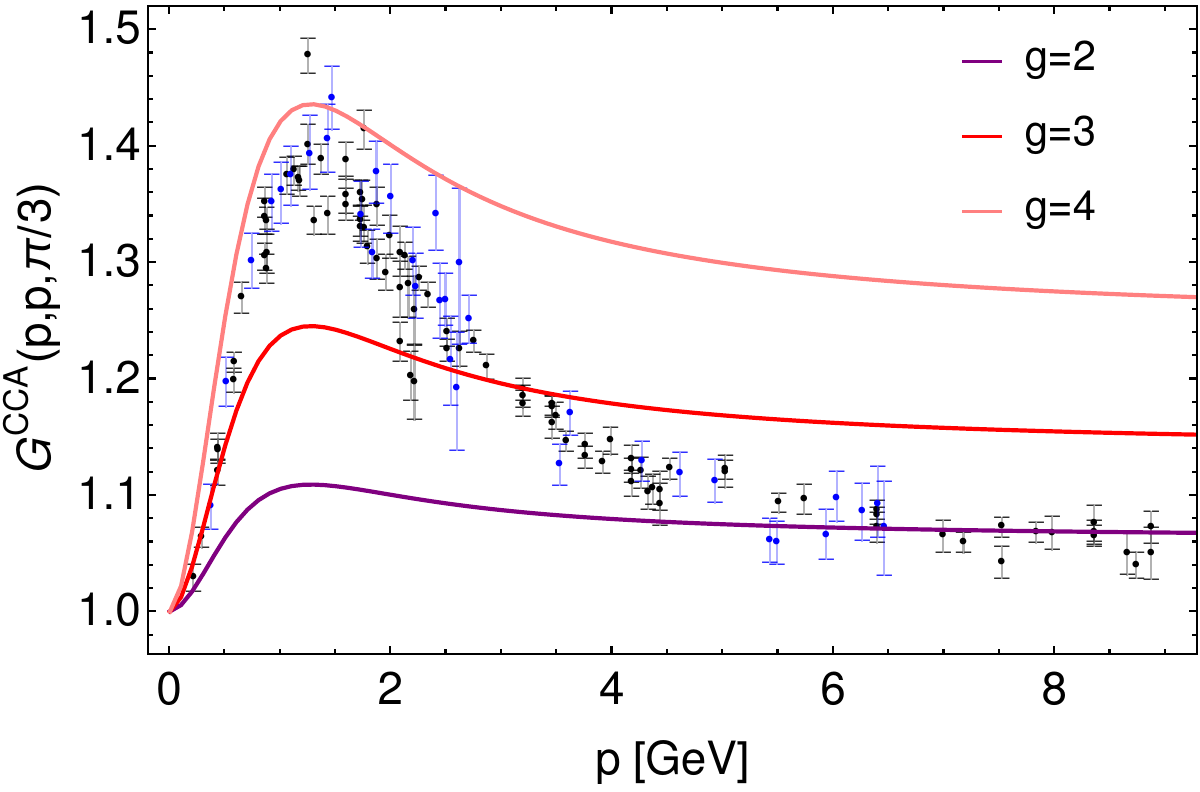}
    \end{subfigure}
    \begin{subfigure}{0.4\textwidth}
        \includegraphics[width=\textwidth]{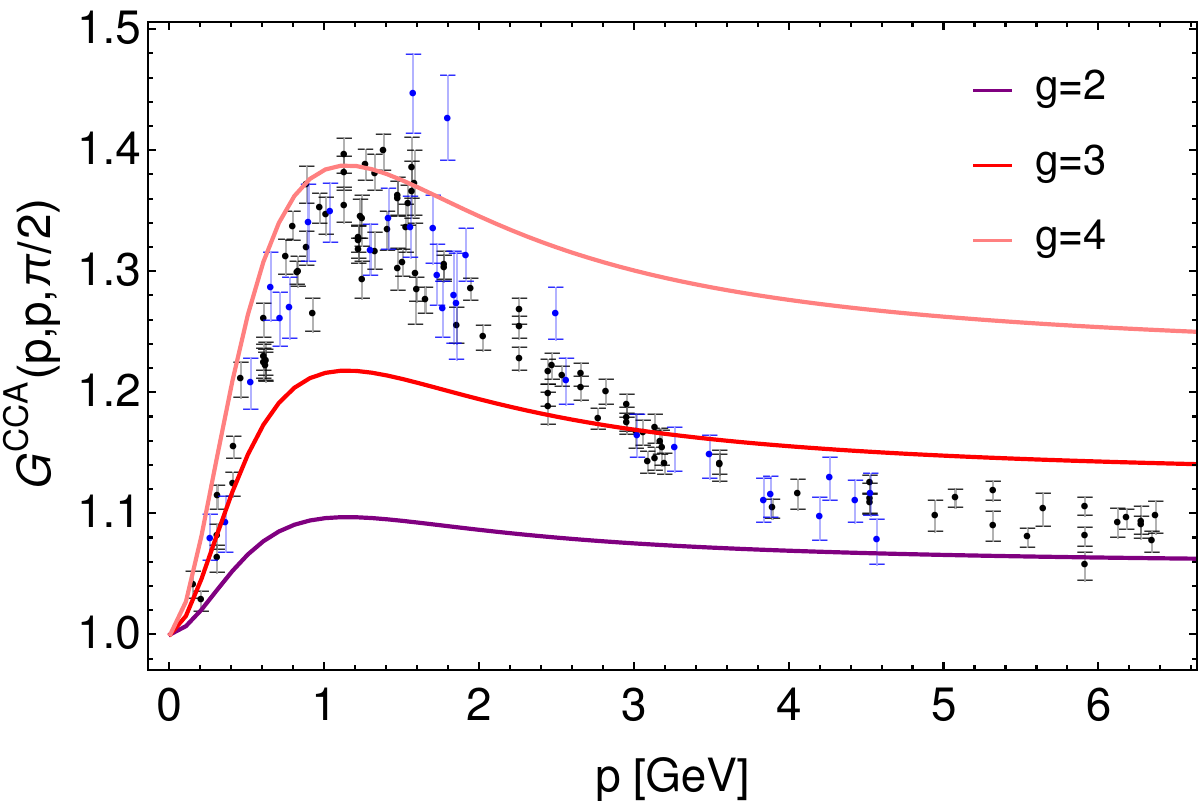}
    \end{subfigure} 
    \begin{subfigure}{0.4\textwidth}
        \includegraphics[width=\textwidth]{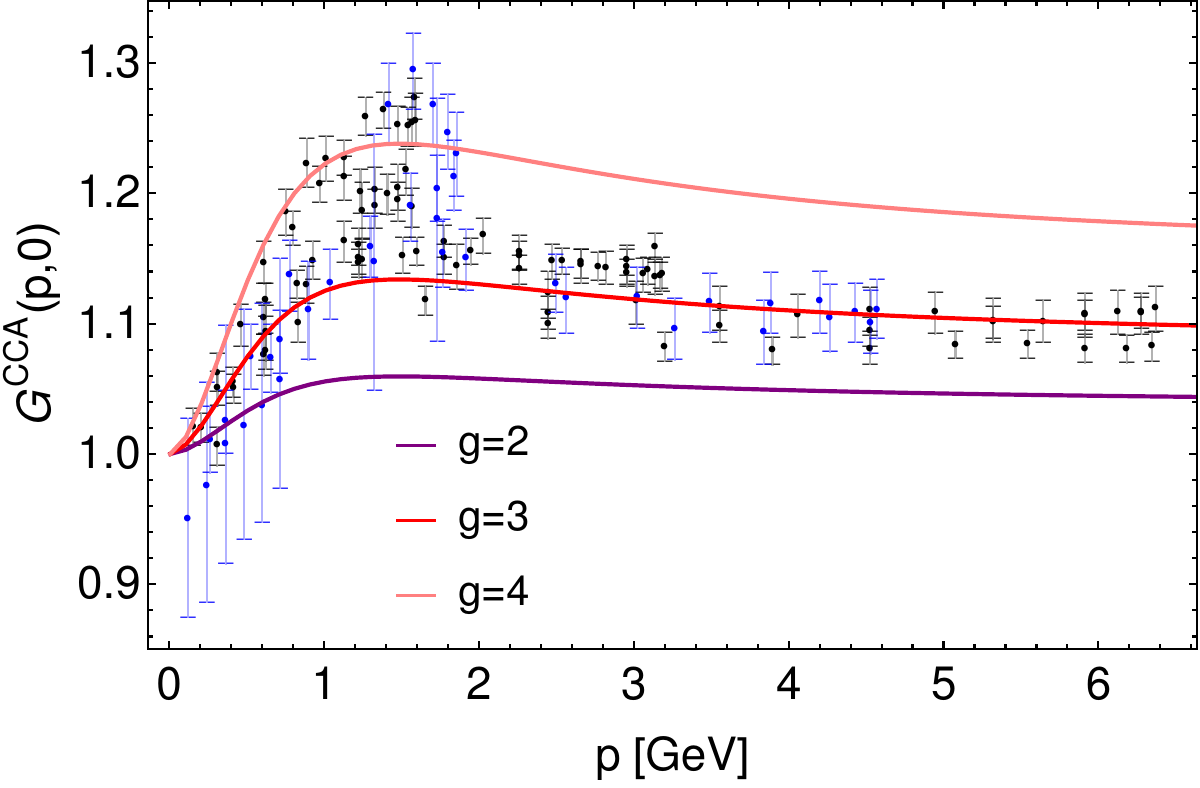}
    \end{subfigure} 
    \caption{The SU(2) scalar function $G^{CCA}$ as a function of the antighost external momentum in three distinct configurations: the symmetric (top left), orthogonal (top right) and the soft gluon limit (bottom) for various values of a fixed gauge coupling. Blue and black points refer to lattice data extracted from refs. \cite{Cucchieri:2008qm} and \cite{Maas:2019ggf}, respectively.}
    \label{fig:SU2_fixed_g}
\end{figure}

In \cref{fig:SU2_fixed_g} we see the dressing function $G^{CCA}$ for various fixed values of the gauge coupling $g$ at one-loop order from the RGZ framework. As for the symmetric and orthogonal kinematic configurations it is clear that, although we find values of the coupling that show a very good agreement with lattice data either at the IR (pink curve) or the UV (purple curve), we are unable to reconcile both regions simultaneously. This is reasonable, as we are not taking into account the RG effects.

Interestingly, in the case of the soft gluon limit it seems that it is possible to describe the lattice data for the whole range of momenta with a constant value of gauge coupling (red curve). The quality of the fit, though, must be taken carefully due to the relatively large dispersion of the lattice data, specially at low momenta.

\begin{figure}[h!]
    \centering
    \begin{subfigure}{0.3\textwidth}
    \includegraphics[width=\textwidth]{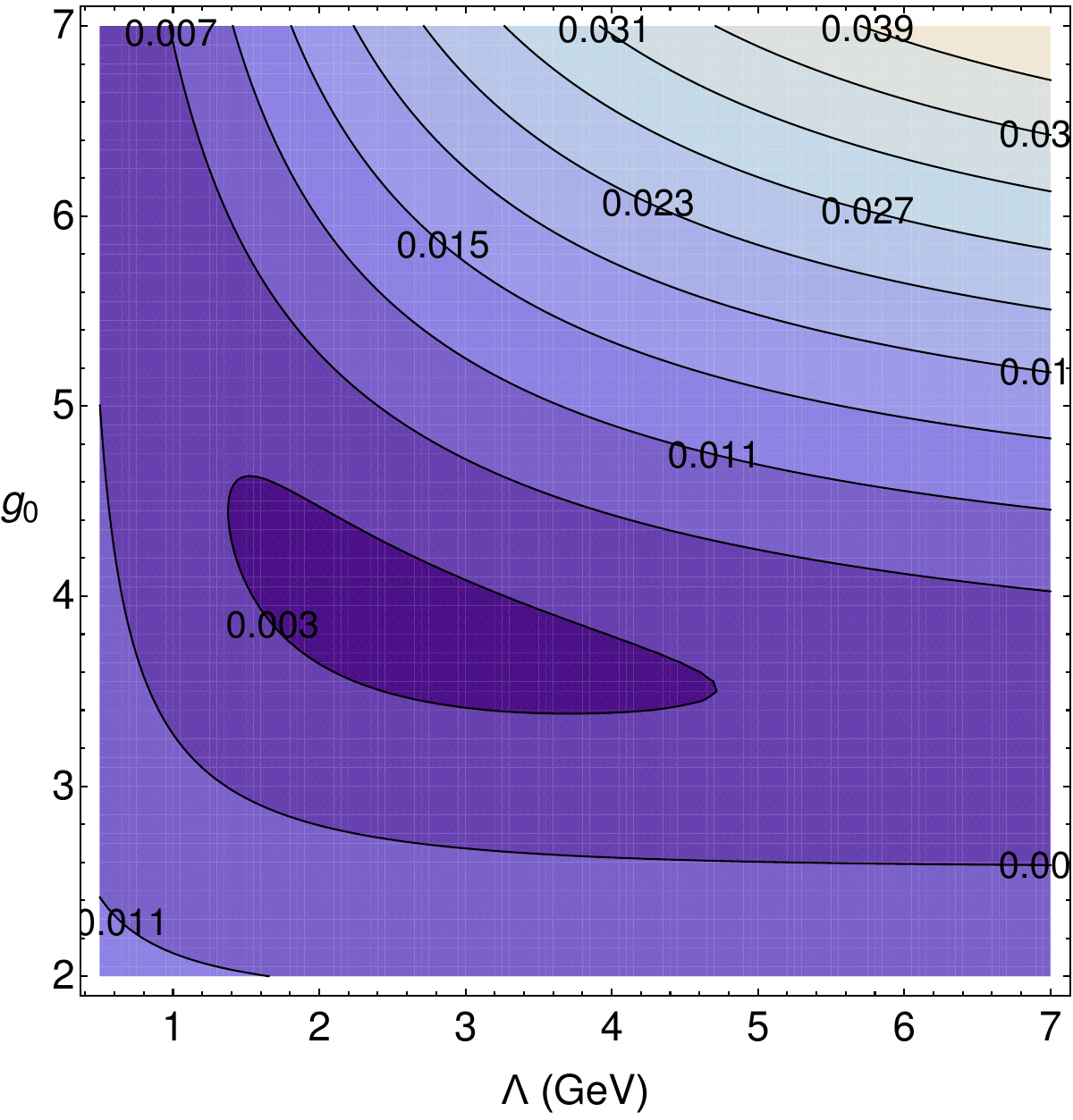}
    \end{subfigure}
    \begin{subfigure}{0.3\textwidth}
        \includegraphics[width=\textwidth]{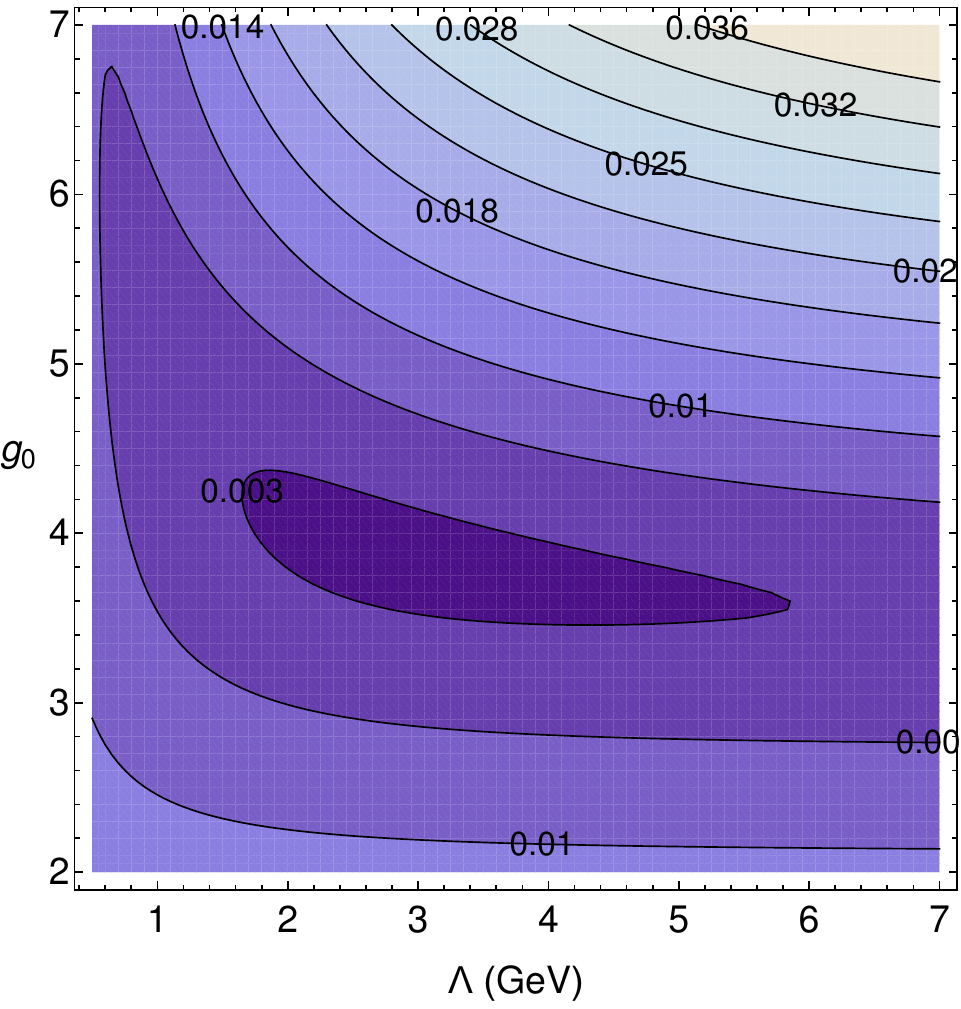}
    \end{subfigure} 
    \begin{subfigure}{0.3\textwidth}
        \includegraphics[width=\textwidth]{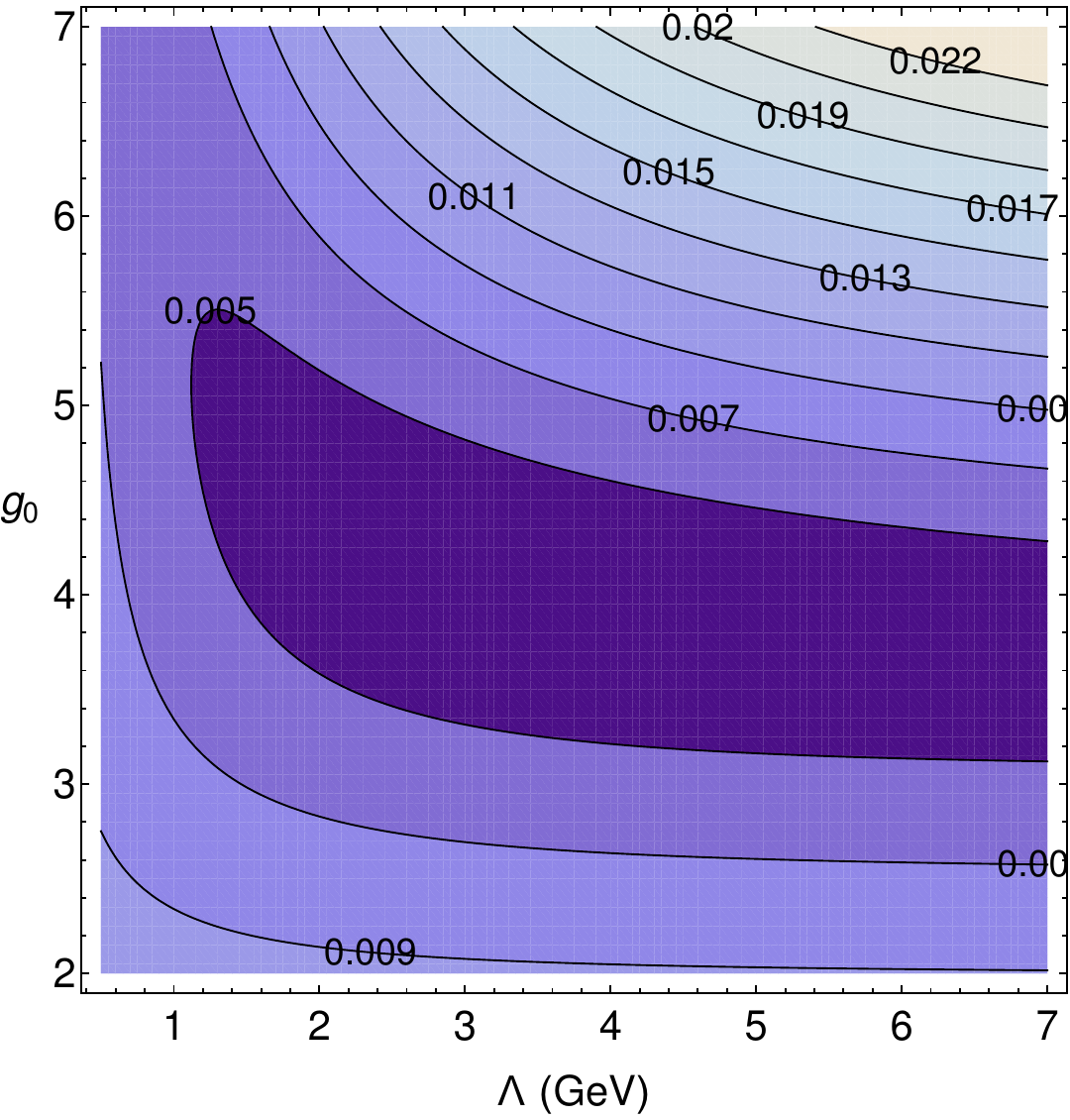}
    \end{subfigure} 
    \caption{Level curves of the relative error $\chi$, defined in \cref{eq:rel_error}, as a function of the parameters $\Lambda$ and $g_0$, for the symmetric (left) and orthogonal (center) configurations as well as the soft gluon limit (right) for the SU(2) case.}
    \label{fig:SU2_level_curves}
\end{figure}

To investigate to which extent the RG flow is capable of reducing the discrepancy between the lattice data and the RGZ fit, we incorporate the toy model of the running coupling shown in eq. (\ref{eq:g_running}). The level curves illustrating the dependence of the relative error defined in \cref{eq:rel_error} on the parameters $g_0$ and $\Lambda$ are depicted in \cref{fig:SU2_level_curves}. We notice that the regions minimizing $\chi$ are very similar for the symmetric and orthogonal configurations. These regions are also compatible with the level curves of $\chi$ observed in the soft gluon limit, though the area of minimum error appears more extensive. This discrepancy could stem from a greater uncertainty and dispersion of the lattice data in the soft gluon limit compared to the data from the symmetric and orthogonal setups.

The parameters that minimize $\chi$ in each kinematic configuration are displayed in \cref{table:parameters_min_chi}. Additionally, we incorporate the values of $g_0$ and $\Lambda$ that minimize the deviation between the lattice data and the RGZ output across all kinematic configurations simultaneously. These values are computed by minimizing the collective error, $\chi_{\text{joint}}$, which we define as

\begin{equation}
 \chi_{\text{joint}}^2\equiv\frac{\chi_{\text{sym.}}^2+\chi_{\text{orth.}}^2+\chi_{\text{soft}}^2}{3}, 
 \label{eq:joint_error}
\end{equation}
where $\chi_{\text{sym.}}$, $\chi_{\text{orth.}}$ and $\chi_{\text{soft}}$ refer to the relative error of the symmetric, orthogonal and the soft gluon configurations, respectively. 

\begin{table}[h!]
  \begin{center} SU(2) fits for different configurations\\ \vspace{.1cm}
    \begin{tabular}{|c||c|c|c|}
    \hline
     Kinematic configuration & \hspace{.2cm} $g_0$ \hspace{.2cm} & \hspace{.2cm} $\Lambda$ (${\rm GeV}$) \hspace{.2cm} & \hspace{.2cm} $\chi$ \hspace{.2cm} \\
\hline \hline
 Symmetric  & 3.90    & 2.40 &   0.0029\\
 \hline
 Orthogonal &  3.85  & 2.90   & 0.0028 \\
 \hline
 Soft gluon & 3.95 & 3.90 & 0.0044 \\
 \hline
 Sym. + Orth. + Soft & 3.90 & 2.75 & 0.0036 \\
 \hline
\end{tabular}  
\end{center} 
    \caption{Values of the parameters $g_0$ and $\Lambda$ that minimize the discrepancy between the RGZ outcome of the function $G^{CCA}$ and the corresponding lattice data for various kinematic configurations in the SU(2) case. The last row refers to the values that minimize the joint error $\chi_{\text{joint}}$, defined in \cref{eq:joint_error}.}
    \label{table:parameters_min_chi}
\end{table}

\begin{figure}[h!]
    \centering
    \begin{subfigure}{0.4\textwidth}
    \includegraphics[width=\textwidth]{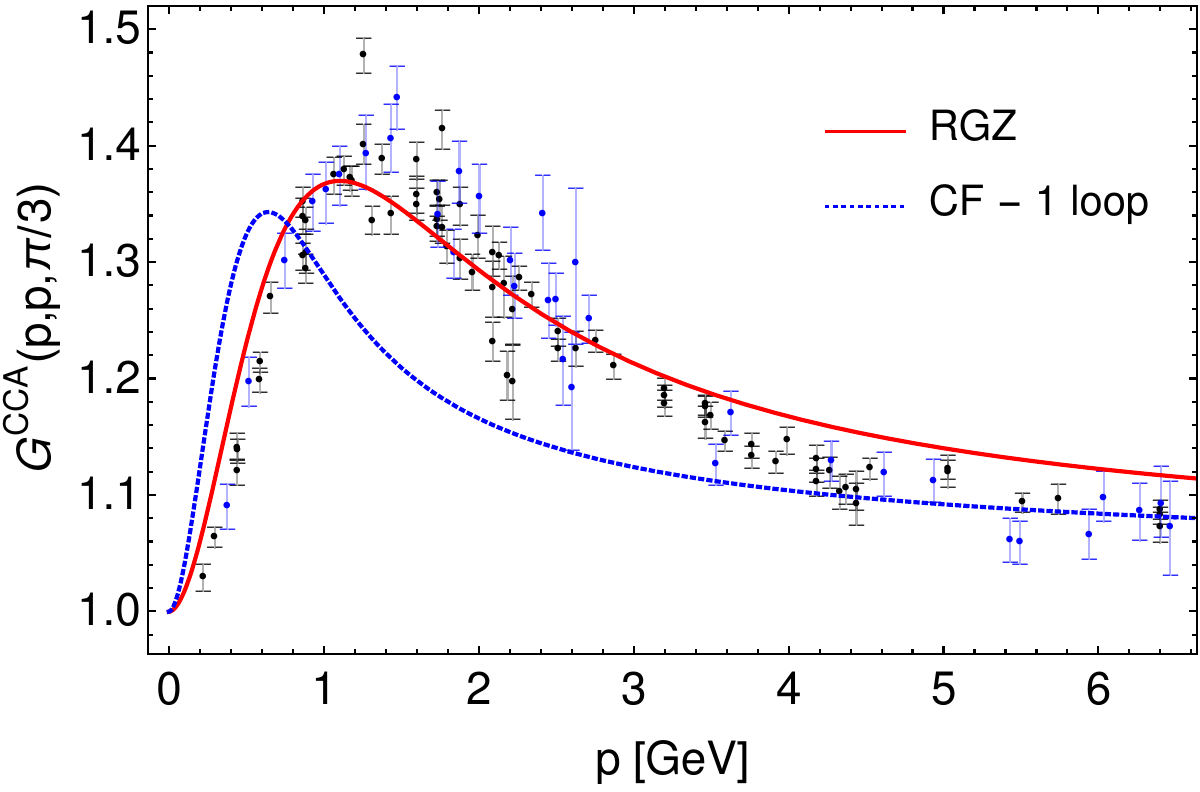}
    \end{subfigure}
    \begin{subfigure}{0.4\textwidth}
        \includegraphics[width=\textwidth]{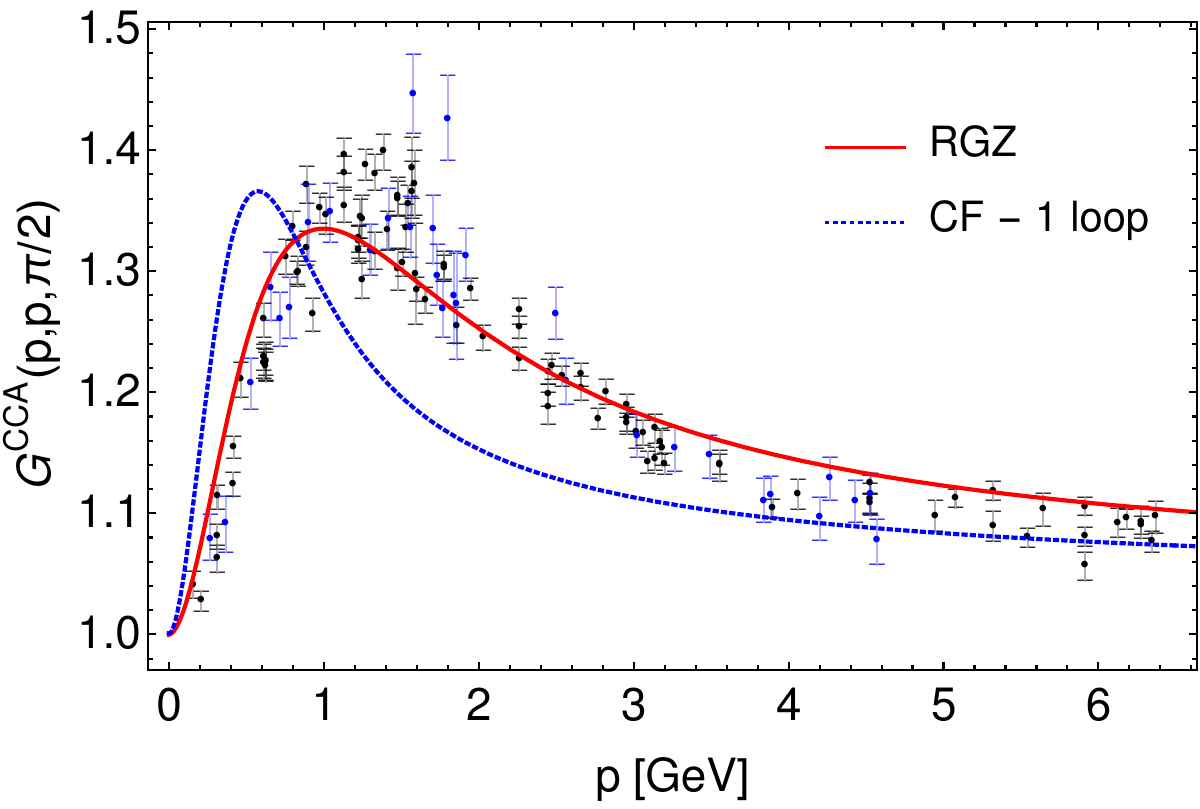}
    \end{subfigure} 
    \begin{subfigure}{0.4\textwidth}
        \includegraphics[width=\textwidth]{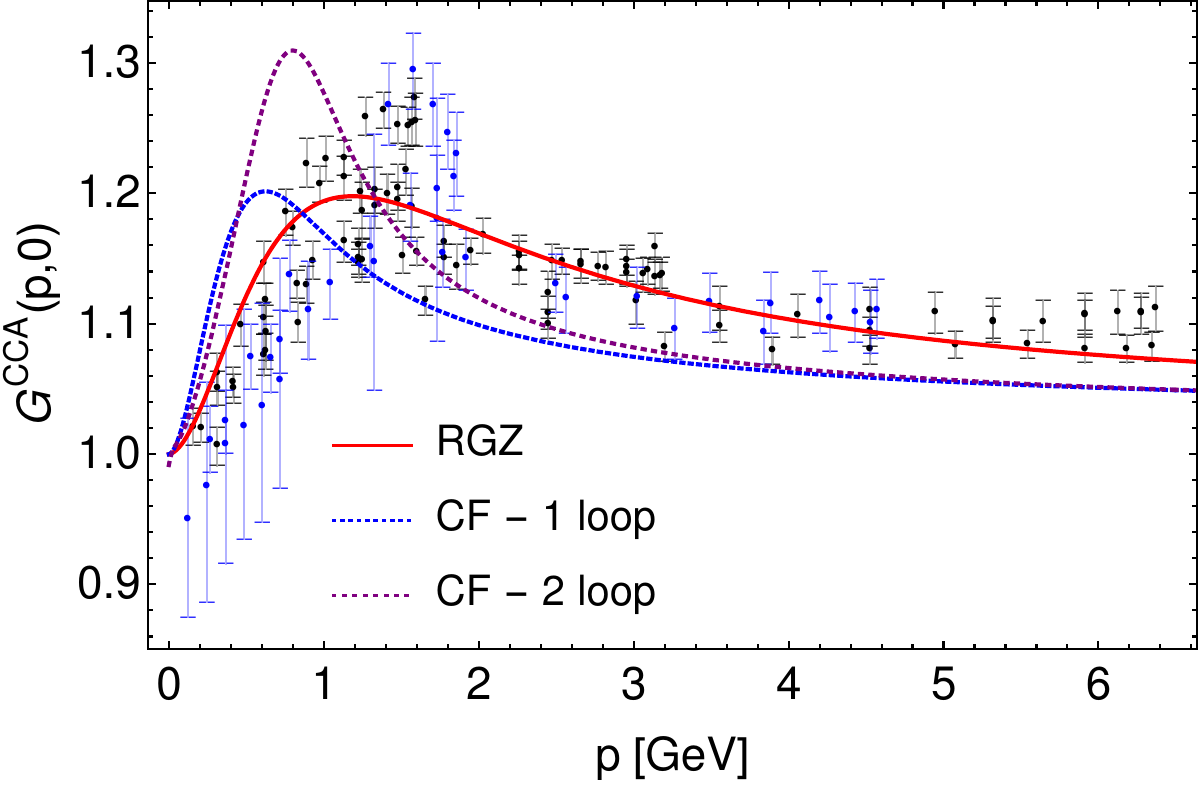}
    \end{subfigure} 
    \caption{The scalar function $G^{CCA}$ for the SU(2) case as a function of the antighost external momentum in three distinct configurations: the symmetric (top left), orthogonal (top right) and the soft gluon limit (bottom). The red curve represents the RGZ fit of the lattice data by using the toy model of \cref{eq:g_running} for the gauge coupling. The values of the parameters, $\Lambda=2.75$ GeV and $g_0=3.9$, are chosen to minimize the joint relative error $\chi_{\text{joint}}$. Regarding the symmetric and orthogonal configurations, the blue curve is the prediction from the CF model at one-loop order \cite{Pelaez:2013cpa}. In the soft gluon limit, the CF results are global fits of the gluon and ghost propagators along with the dressing function $G^{CCA}$ at one- and two-loop order \cite{Barrios:2020ubx}. Blue and black points refer to lattice data extracted from refs. \cite{Cucchieri:2008qm} and \cite{Maas:2019ggf}, respectively.}
    \label{fig:SU2_fit_RG}
\end{figure}

The parameters that minimize $\chi_{\text{joint}}$ are $g_0=3.90$ and $\Lambda=2.75\ {\rm GeV} $. Using these parameter values, we collect in \cref{table:chi_confs_from_min_chi_joint} the $\chi$ values for each kinematic configuration, which indicate that the deterioration of the individual fits for each configuration is minimal when performing the collective fit. The corresponding plots are shown in \cref{fig:SU2_fit_RG}.

\begin{table}[h!]
  \begin{center} SU(2) joint fit\\ \vspace{.1cm}
    \begin{tabular}{|c||c|}
    \hline
     Kinematic configuration & \hspace{.2cm} $\chi$ \hspace{.2cm} \\
\hline \hline
 Symmetric  & 0.0031 \\
 \hline
 Orthogonal & 0.0028 \\
 \hline
 Soft gluon & 0.0046 \\
 \hline
\end{tabular}  
\end{center}
    \caption{Values of the parameters $\chi$ of each kinematic configuration corresponding to the parameters that minimize the joint error $\chi_{\text{joint}}$, $g_0=3.90$ and $\Lambda=2.75 {\rm GeV} $, in the SU(2) case.}
    \label{table:chi_confs_from_min_chi_joint}
\end{table}

We observe that the fit of \cref{fig:SU2_fit_RG} exhibits a strong agreement with lattice simulations across all kinematic configuration. However, the UV tails of the RGZ curves consistently fall above the lattice data points. This discrepancy could arise from the larger number of lattice points in the IR region compared to the UV domain, emphasizing the agreement between the RGZ result and lattice data in the low momentum zone at the expense of deteriorating the fit in the UV.

In \cref{fig:SU2_fit_RG} we include also the results of the Curci-Ferrari (CF) model in Landau gauge at one- and two-loop accuracy. The outcomes from both frameworks, CF and RGZ, are consistent. Nonetheless, within this analysis, comparing both approaches must be done with care due to two reasons. Firstly, concerning the symmetric and orthogonal configurations, the CF curves of $G^{CCA}$ are pure predictions of the model, while the RGZ curves represent a global fit of the lattice data. This renders the comparison fairer in the case of the soft-gluon limit, where the results from the CF approach come from a global fit of the two-point functions and $G^{CCA}$. Secondly, due to the complexity of the RGZ approach, we employ a simplified model for the RG flow of the gauge coupling, disregarding the RG effects of other parameters of the theory, which could potentially alter the final result. 

\subsubsection{Contribution of each diagram to the $G^{CCA}$ fit}

\begin{figure}[h!]
    \centering
    \begin{subfigure}{0.4\textwidth}
    \includegraphics[width=\textwidth]{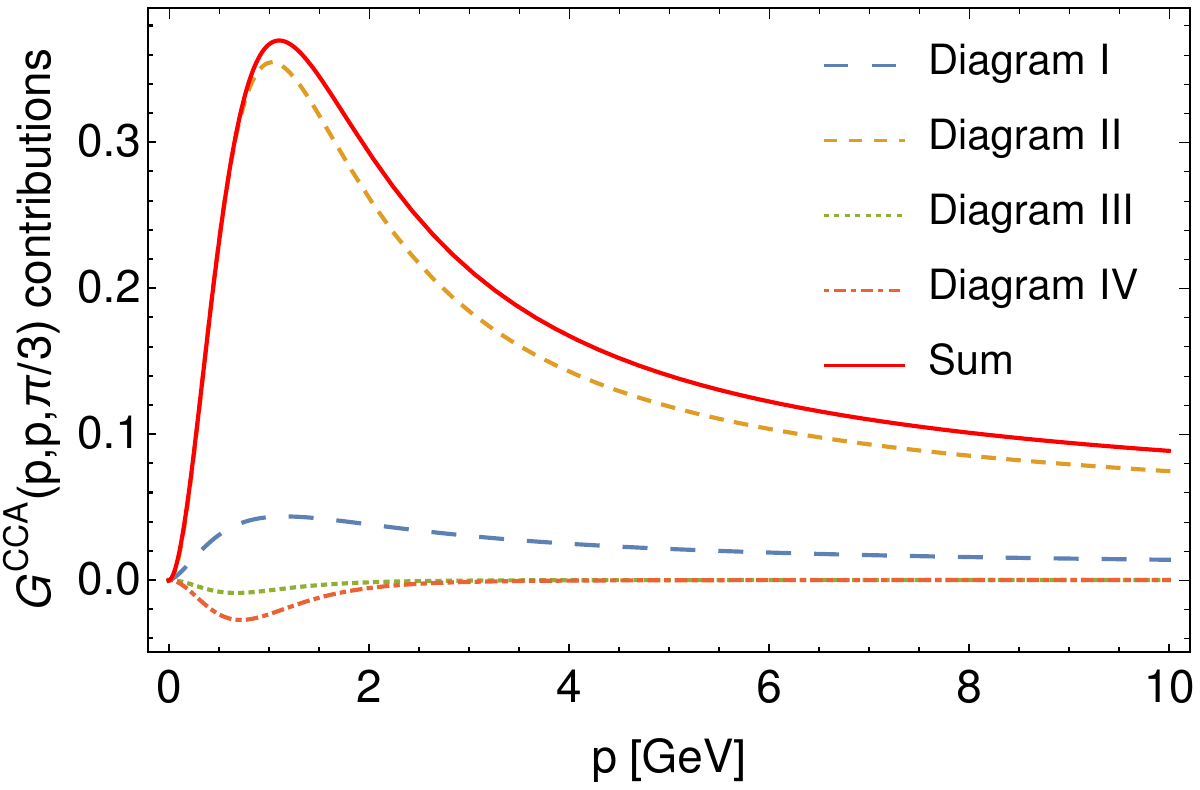}
    \end{subfigure}
    \begin{subfigure}{0.4\textwidth}
        \includegraphics[width=\textwidth]{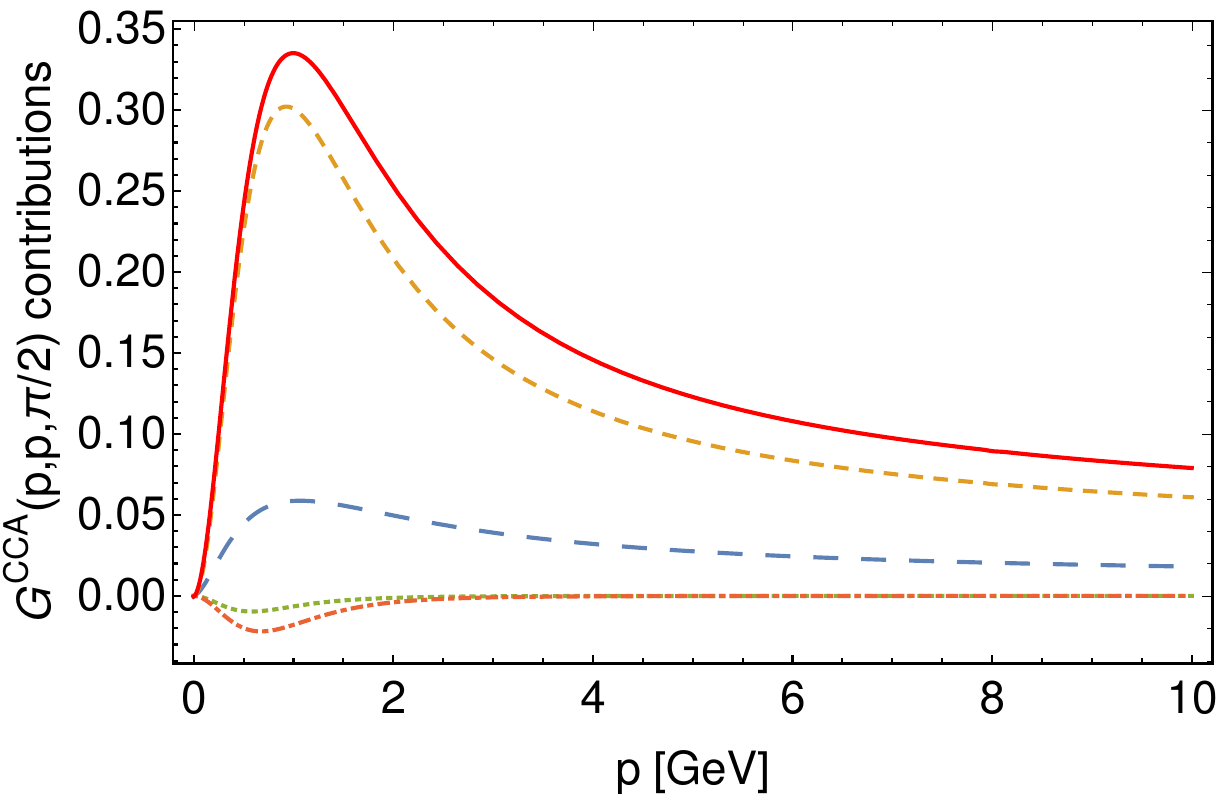}
    \end{subfigure} 
    \begin{subfigure}{0.4\textwidth}
        \includegraphics[width=\textwidth]{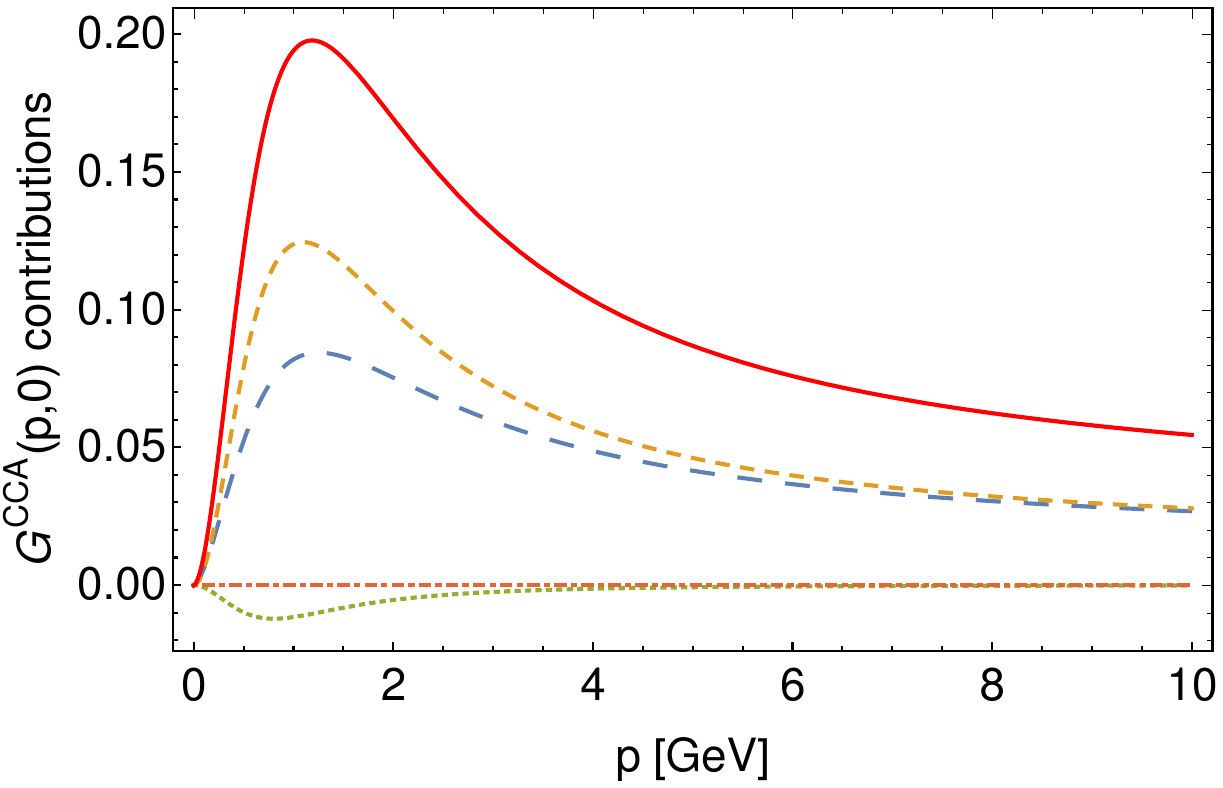}
    \end{subfigure} 
    \caption{Contribution of each diagram to $G^{CCA}$ for the SU(2) case in three distinct configurations: the symmetric (top left), orthogonal (top right) and the soft gluon limit (bottom), using the gauge coupling described by \cref{eq:g_running}. The values of the parameters, $\Lambda=2.75$ GeV and $g_0=3.90$, are chosen to minimize the joint relative error $\chi_{\text{joint}}$. The convention of colors and line-styles in the orthogonal and soft configurations is the same as the symmetric case. The red curve is the sum of all diagrams. The numbering of the diagrams is provided in \cref{fig:feyndiags}. Diagram (IV) refers to the full contribution $\frac{2g\gamma^2f^{ade}}{k^2+M^2}\Gamma_{\bar c^b c^c \varphi^{de}_\mu }(k,p,r)$.}
    \label{fig:SU2_contribution_per_diag_GCCA}
\end{figure}

We now examine the contribution of each diagram to the full result of $G^{CCA}$ in the symmetric and orthogonal configurations, as well as in the soft-gluon limit. The result is presented in \cref{fig:SU2_contribution_per_diag_GCCA}, where we follow the numbering from \cref{fig:feyndiags}. As for the quantity we denote as diagram (IV), it corresponds to the full contribution $\frac{2g\gamma^2f^{ade}}{k^2+M^2}\Gamma_{\bar c^b c^c \varphi^{de}_\mu }(k,p,r)$, see \cref{eq:Acc-local-shorthand-notation}. 

We note that genuinely RGZ diagrams, (III) and (IV), represent a small contribution at one-loop order as compared to the largest contribution of diagram (II) and, to a lesser degree, diagram (I). This is in agreement with the results obtained by CF model \cite{Tissier:2010ts,Tissier:2011ey,Pelaez:2013cpa,Pelaez:2014mxa,Pelaez:2015tba,Gracey:2019xom,Pelaez:2021tpq,Figueroa:2021sjm,Barrios:2020ubx,Barrios:2021cks,Barrios:2022hzr} where only QCD-like diagrams with massive gluons are taken into account. Interestingly, as for the symmetric and orthogonal configurations, the effect of diagrams (III) and (IV) is a reduction of the peak of $G^{CCA}$, at around $p=1.5$ GeV. In the case of the soft gluon limit, we observe that the contributions coming from diagrams (I) and (II) are of similar order whereas the contribution of diagram (IV) vanishes, as already pointed out in \cref{sec:soft-gluon}.

\subsection{SU(3)}

In this section we replicate the analysis performed for the SU(2) gauge group but for the case of the SU(3) gauge group. Yet, the conclusions derived from this analysis will be more restricted compared to the SU(2) case due to the lack of lattice data available in kinematic configurations other than the soft gluon scenario. 

In what follows we present  results for the same configurations analyzed in the previous section, \textit{i.e.} the symmetric and orthogonal configurations as well as in the soft gluon limit. Although the only configuration with available lattice data is the latter, the other configurations are valuable for comparison with alternative approaches in the continuum.   

\subsubsection{Fitting the ghost-gluon vertex dressing function}

\begin{figure}[h!]
    \centering
    \begin{subfigure}{0.4\textwidth}
    \includegraphics[width=\textwidth]{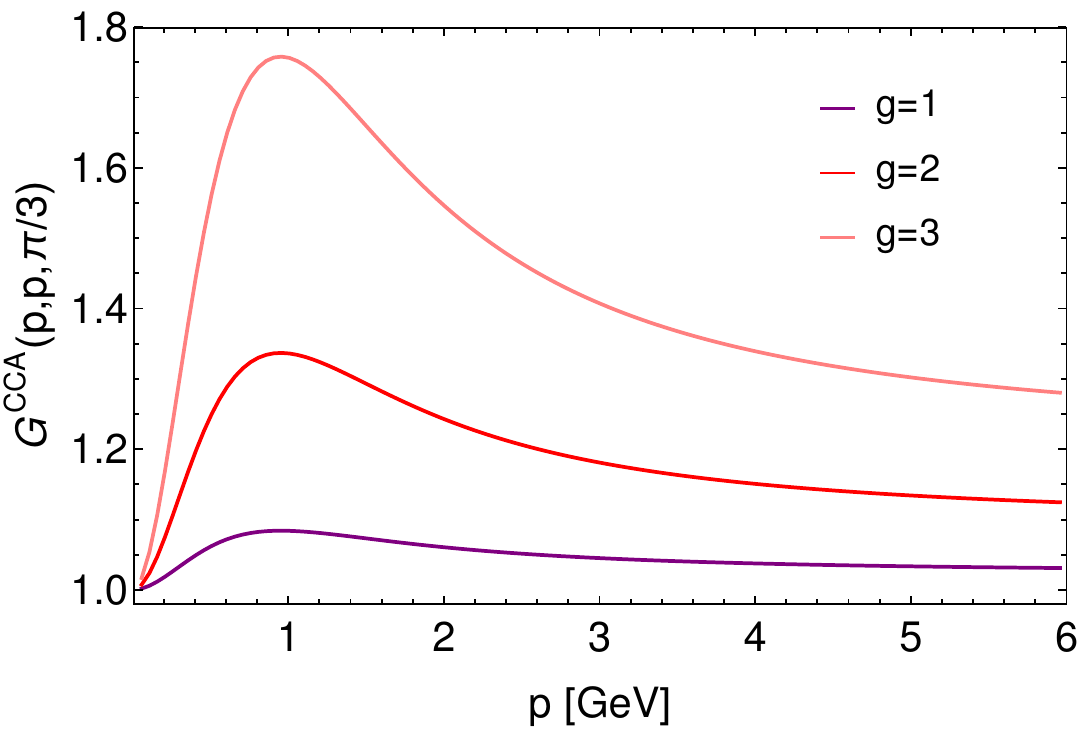}
    \end{subfigure}
    \begin{subfigure}{0.4\textwidth}
        \includegraphics[width=\textwidth]{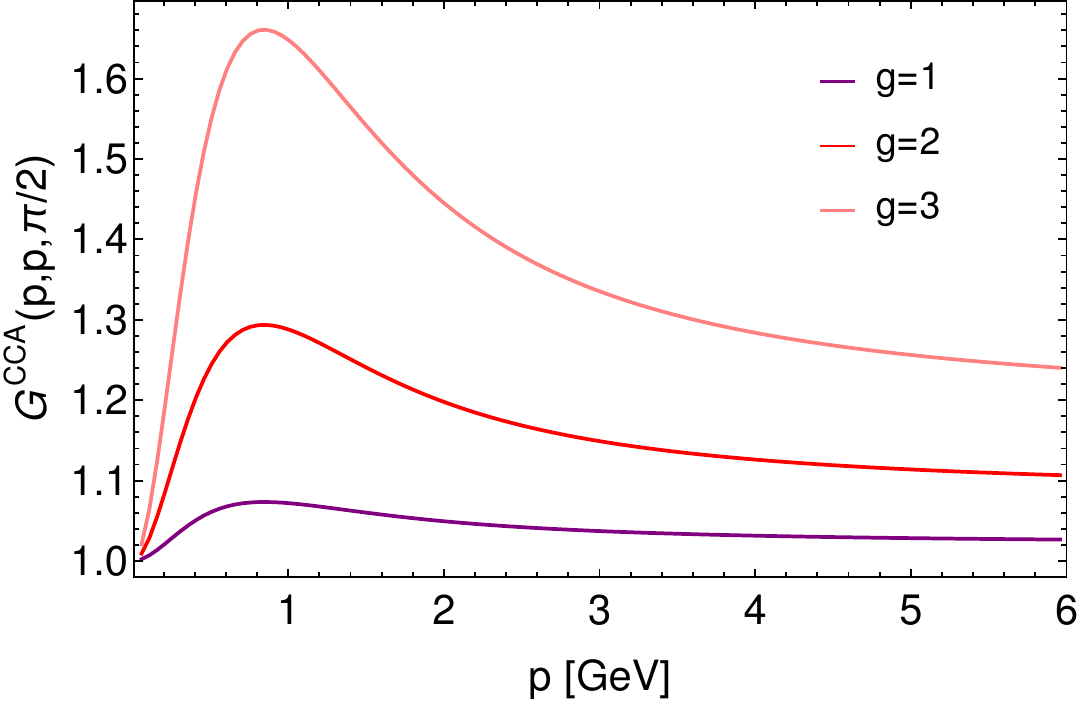}
    \end{subfigure} 
    \begin{subfigure}{0.4\textwidth}
        \includegraphics[width=\textwidth]{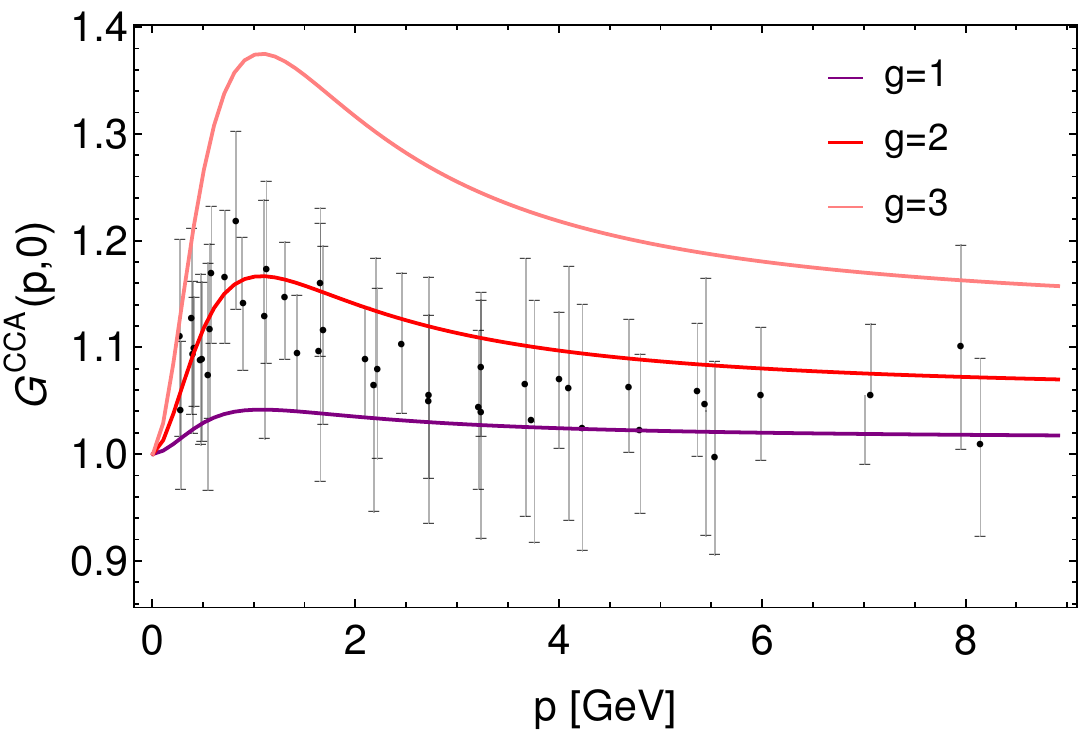}
    \end{subfigure} 
    \caption{The scalar function $G^{CCA}$ for the SU(3) case as a function of the antighost external momentum in three distinct configurations: the symmetric (top left), orthogonal (top right) and the soft gluon limit (bottom) for various values of a fixed gauge coupling. The lattice data were extracted from the plots in \cite{Ilgenfritz:2006he} using WebPlotDigitizer \cite{webplotdigitizer}. We estimated the error of the extraction procedure to be at most 0.8\%.}.
    \label{fig:SU3_fixed_g}
\end{figure}

In \cref{fig:SU3_fixed_g} we see $G^{CCA}$ as a function of the antighost momentum for the symmetric and orthogonal configurations as well as in the soft gluon limit. Similarly to what we observed in the case of the SU(2) gauge group, as for the soft gluon limit it seems that we are capable of reproducing the lattice data by using a fixed value of the gauge coupling $g$, as was already pointed out in \cite{Mintz:2017qri}. 

\begin{figure}[h!]
    \centering
    \includegraphics[width=.3\textwidth]{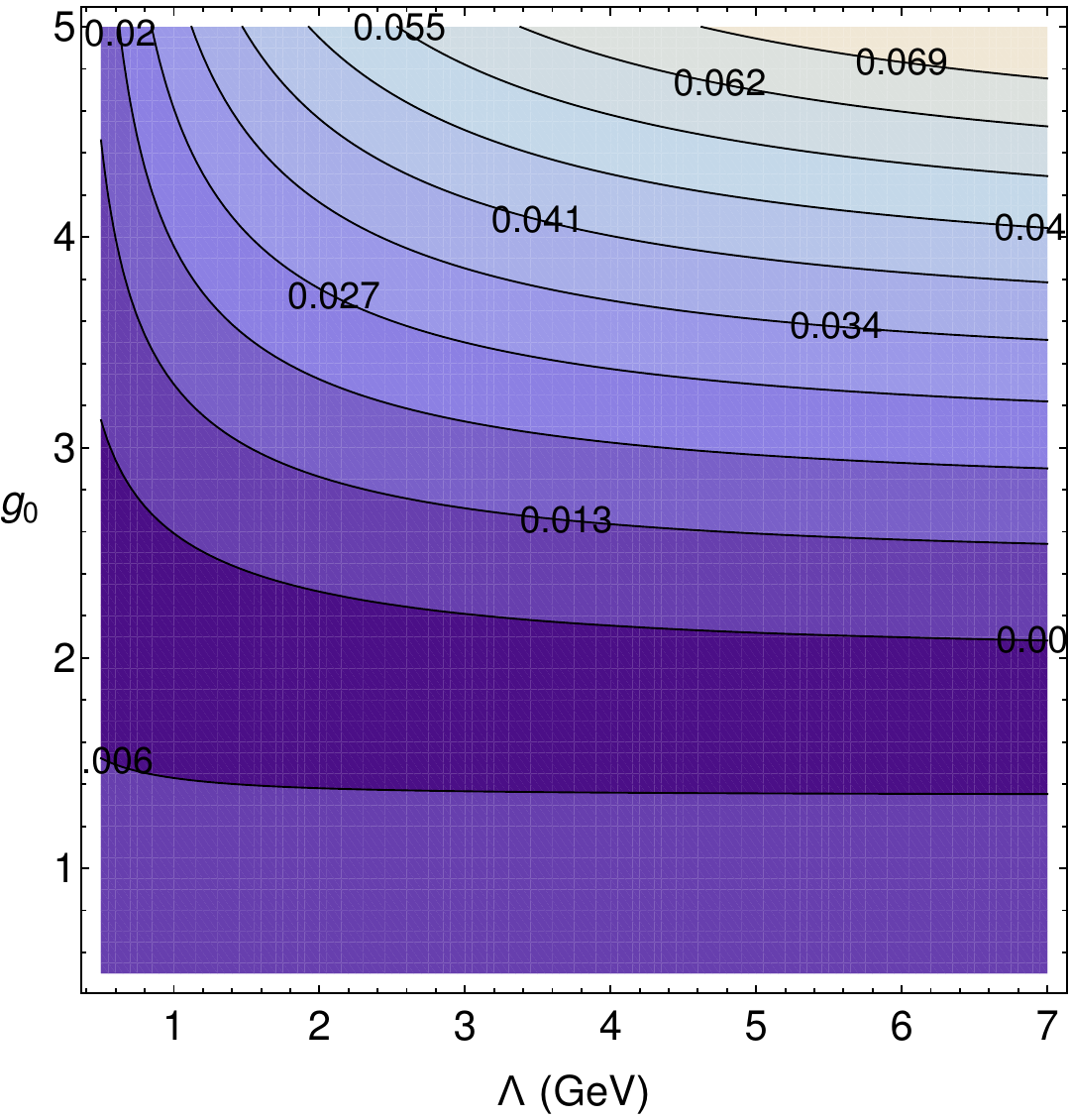}
    \caption{Level curves of the relative error $\chi$, defined in \cref{eq:rel_error}, as a function of the parameters $\Lambda$ and $g_0$, as for the soft gluon limit in the SU(3) case.}.
    \label{fig:SU3_soft_level_curves}
\end{figure}

It is worth examining how the deviation between the RGZ outcome and lattice data is modified as we change the parameters $\Lambda$ and $g_0$ when using the simplified model (eq. (\ref{eq:g_running})). The level curves illustrating this dependence are depicted in \cref{fig:SU3_soft_level_curves}, displaying an error $\chi$ independent of the scale $\Lambda$ as long as the values of $g_0$ remain moderate. This reinforces our previous observation that, within the soft gluon limit, the RGZ framework effectively reproduces lattice data with a fixed value of $g$. Nevertheless, as we aim to analyze the symmetric and orthogonal configurations as well, which are likely sensitive to RG effects, in what follows we continue employing the scale-dependent gauge coupling defined by \cref{eq:g_running}.

\begin{figure}[h!]
    \centering
    \begin{subfigure}{0.4\textwidth}
    \includegraphics[width=\textwidth]{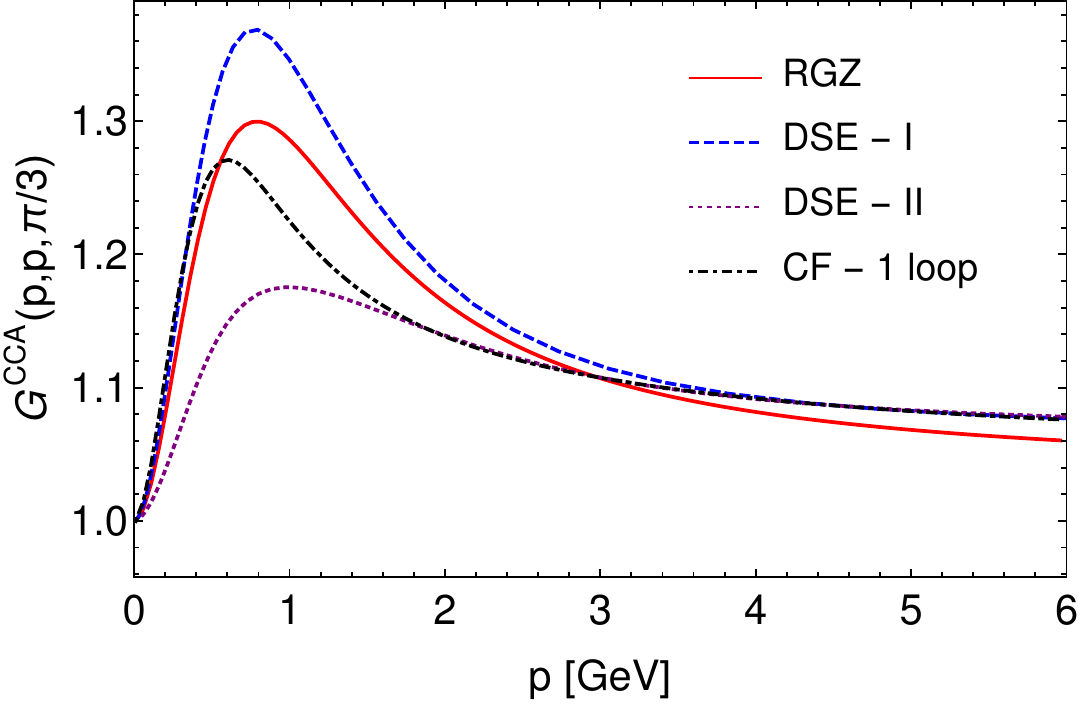}
    \end{subfigure}
    \begin{subfigure}{0.4\textwidth}
        \includegraphics[width=\textwidth]{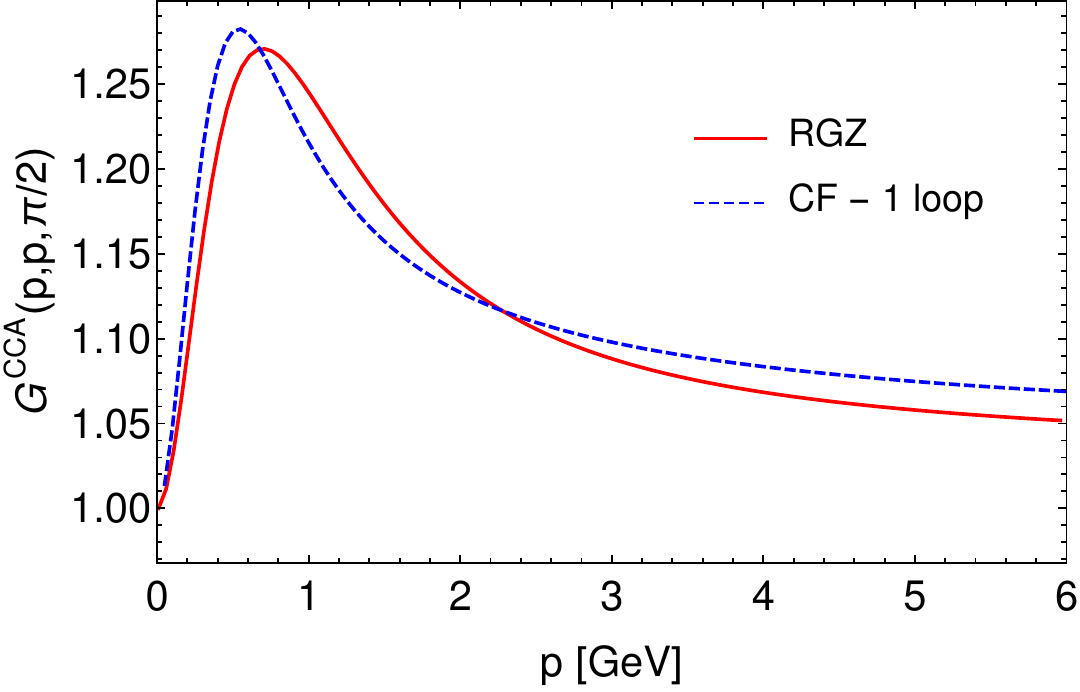}
    \end{subfigure} 
    \begin{subfigure}{0.4\textwidth}
        \includegraphics[width=\textwidth]{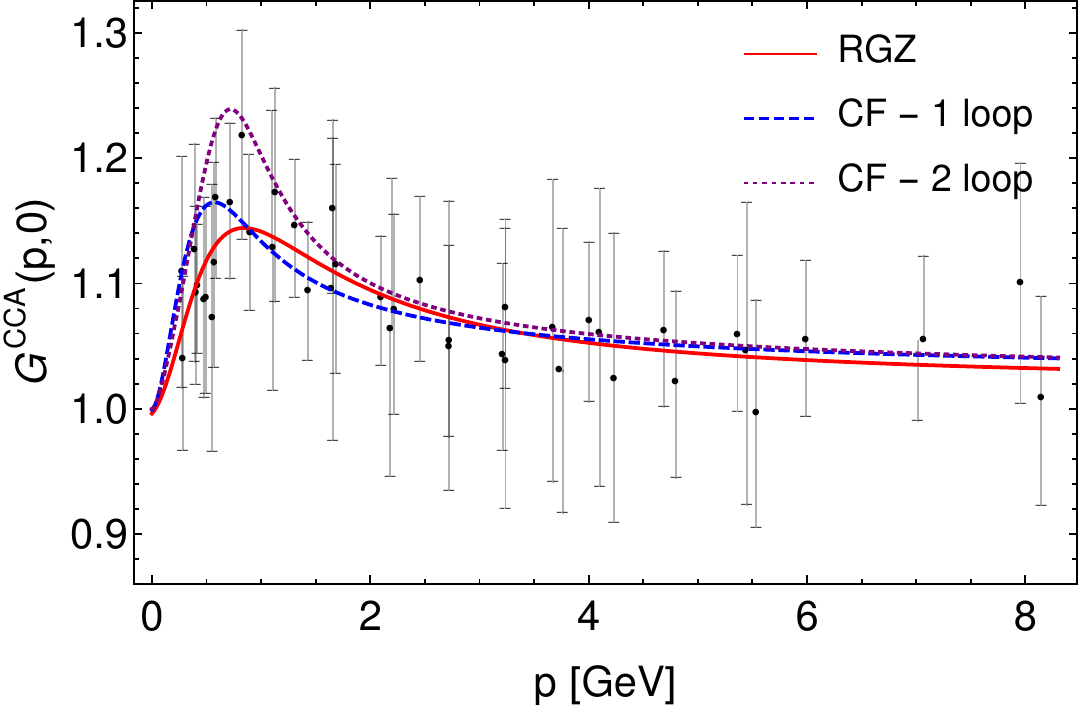}
    \end{subfigure} 
    \caption{The scalar function $G^{CCA}$ for the SU(3) case as a function of the antighost external momentum in three distinct configurations: the symmetric (top left), orthogonal (top right) and the soft gluon limit (bottom). The red curve corresponds to the RGZ fit of the lattice data, by utilizing the toy model of \cref{eq:g_running} for the gauge coupling. Parameters $\Lambda=0.70$ GeV and $g_0=2.15$ were chosen to minimize the joint relative error with the lattice data in the case of the soft gluon limit. Lattice data are extracted from \cite{Ilgenfritz:2006he}. We compare the results with other approaches in the continuum. Specifically with $G^{CCA}$ obtained from DSE, considering the bare (DSE - I) and the dressed three-gluon vertex (DSE - II) \cite{Aguilar:2018csq} and $G^{CCA}$ from the CF model at one- and two-loop order \cite{Pelaez:2013cpa,Barrios:2020ubx}.}
    \label{fig:SU3_fit_RG}
\end{figure}

In analogy to our analysis for the SU(2) case, in order to find the best fit to lattice data, we find the parameters $\Lambda$ and $g_0$ that minimize the relative error $\chi$. In the soft gluon setup, the minimum of $\chi$  corresponds to $\Lambda=0.70$ GeV and $g_0=2.15$, with a relative error of $\chi=0.0040$. The resulting plots are depicted in \cref{fig:SU3_fit_RG}. 

We note that our results for $G^{CCA}$ are consistent with the outcomes of the CF model \cite{Pelaez:2013cpa,Barrios:2020ubx} and DSE \cite{Aguilar:2018csq}. Furthermore, the RGZ fit shows a strong agreement with simulations, resembling the results obtained with a fixed value of the gauge coupling. Interestingly, while the UV tails of the DSE and CF curves overlap in the UV, the UV tails from the RGZ curves fall more rapidly in comparison.

\subsubsection{Contribution of each diagram to the $G^{CCA}$ fit}

\begin{figure}[h!]
    \centering
    \begin{subfigure}{0.4\textwidth}
    \includegraphics[width=\textwidth]{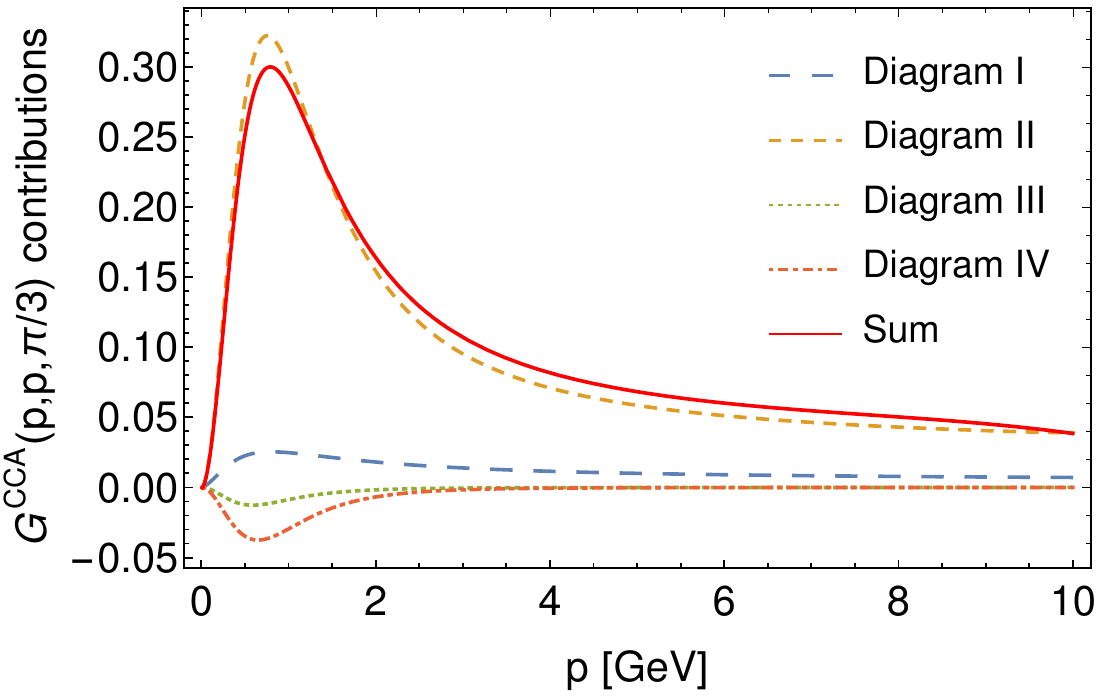}
    \end{subfigure}
    \begin{subfigure}{0.4\textwidth}
        \includegraphics[width=\textwidth]{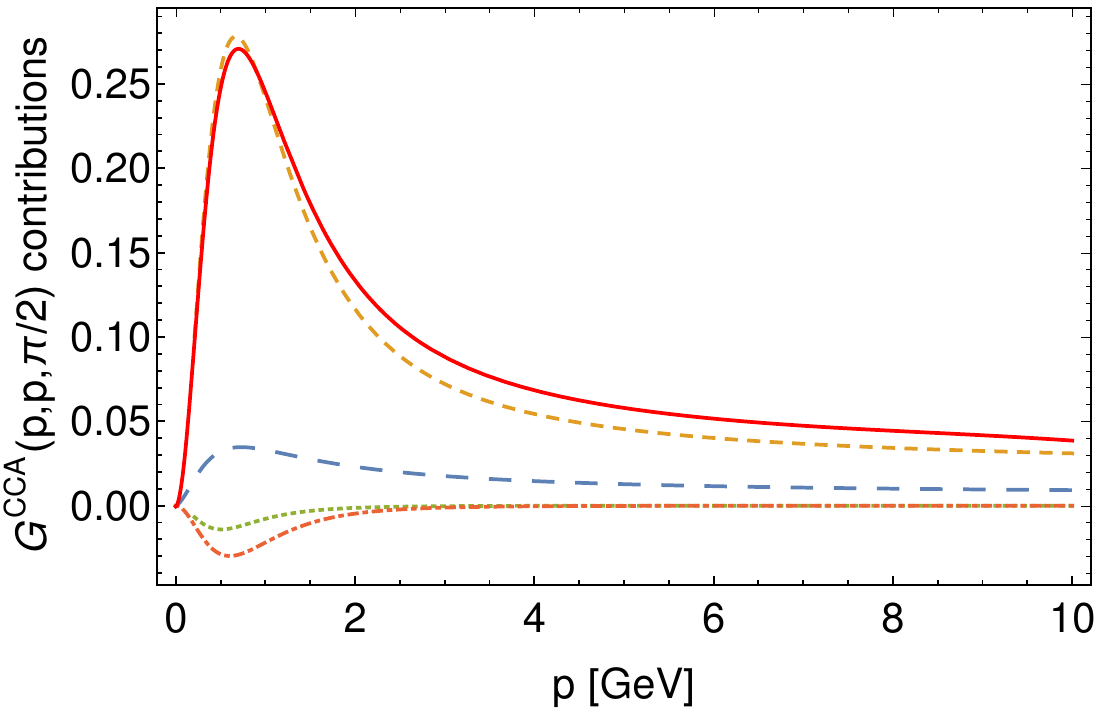}
    \end{subfigure} 
    \begin{subfigure}{0.4\textwidth}
        \includegraphics[width=\textwidth]{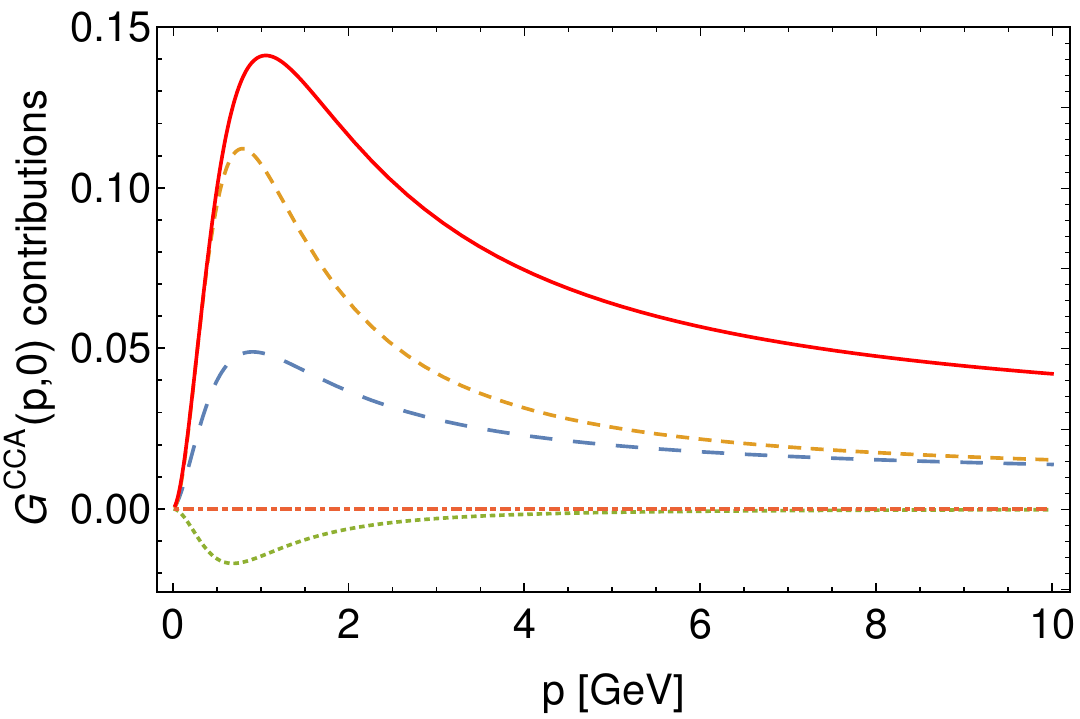}
    \end{subfigure} 
    \caption{Contribution of each diagram to $G^{CCA}$ for the SU(3) case in three distinct configurations: the symmetric (top left), orthogonal (top right) and the soft gluon limit (bottom), using the gauge coupling described by \cref{eq:g_running}. Parameters, $\Lambda=0.70$ GeV and $g_0=2.15$, are chosen to minimize the relative error $\chi$ in the soft gluon limit. The convention of colors and line-styles in the orthogonal and soft configurations is the same as the symmetric case. The solid, red curve is the sum of all diagrams. The numbering of the diagrams is provided in \cref{fig:feyndiags}. Diagram (IV) refers to the full contribution $\frac{2g\gamma^2f^{ade}}{k^2+M^2}\Gamma_{\bar c^b c^c \varphi^{de}_\mu }(k,p,r)$.}
    \label{fig:SU3_contribution_per_diag_GCCA}
\end{figure}

To conclude this section, we investigate the contribution of each diagram to ghost-gluon vertex dressing function, which is illustrated in \cref{fig:SU3_contribution_per_diag_GCCA}. Proceeding similarly to the SU(2) case, the quantity we call diagram (IV) corresponds to the full contribution $\frac{2g\gamma^2f^{ade}}{k^2+M^2}\Gamma_{\bar c^b c^c \varphi^{de}_\mu }(k,p,r)$, see \cref{eq:Acc-local-shorthand-notation}.

We observe that in the case of the symmetric and orthogonal configurations the behaviour of $G^{CCA}$ is dominated by diagram (II), while the contribution of diagram (I) is counterbalanced by the contributions of the genuine RGZ diagrams (III) and (IV). In the case of the soft gluon limit, diagrams (I) and (II) both significantly contribute to the function $G^{CCA}$, whereas diagram (IV) vanishes, as explained in \cref{sec:soft-gluon}. Overall, the relative weights of these contributions do not significantly differ from the SU(2) case.

\section{Influence of the toy model on the UV behavior of the function $G^{CCA}$}
\label{sec:g_toy_UV}

As mentioned earlier, when comparing the UV tails of the dressing function $G^{CCA}$ obtained from the RGZ framework with other approaches, cf.  \cref{fig:SU2_fit_RG,fig:SU3_fit_RG}, we observe differences. Particularly striking is the observation in the case of SU(3), where a simultaneous comparison between the RGZ, the CF model and DSE results reveals a convergence in the tails from the latter approaches, starting from $p=4$ GeV onwards. In contrast, the UV tails of the curves from the RGZ framework decrease more rapidly. It is legitimate then to ask for the potential causes of this discrepancy in the UV. In this section we explore the extent to which this difference may derive from the toy model of the RG flow, as presented in \cref{eq:g_running}, and its associated parameters $\Lambda$ and $g_0$.

\begin{figure}[h!]
    \centering
    \begin{subfigure}{0.4\textwidth}
    \includegraphics[width=\textwidth]{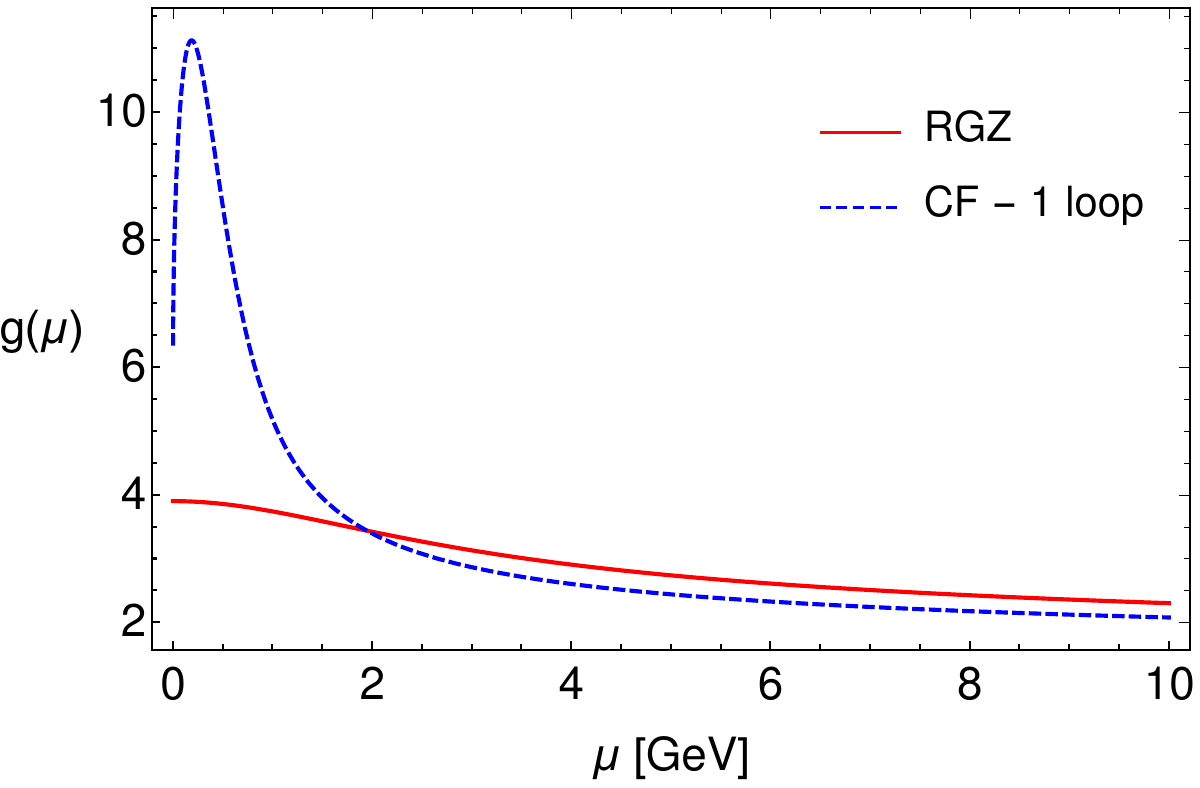}
    \end{subfigure}
    \begin{subfigure}{0.4\textwidth}
        \includegraphics[width=\textwidth]{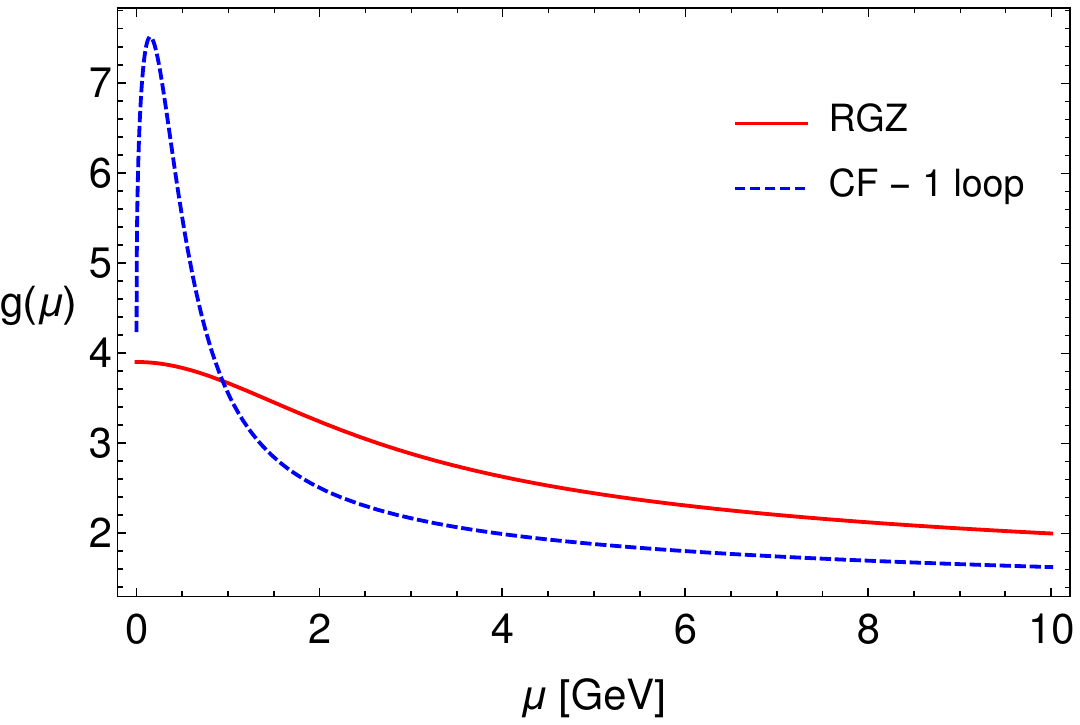}
    \end{subfigure}
    \caption{The running coupling $g_{\text{toy}}$ and $g_{\text{CF}}$ for the SU(2) (left) and SU(3) (right) gauge groups. As for the toy model, the values of the parameters are $\Lambda=2.4,$  GeV, $g_0=3.9$ and $\Lambda=0.70$  GeV, $g_0=2.15$, for SU(2) and SU(3), respectively.}
    \label{fig:g_comparison}
\end{figure}

With that purpose we compare in \cref{fig:g_comparison} the running of the gauge coupling employed for the CF prediction of $G^{CCA}$ at one-loop order in \cref{fig:SU2_fit_RG,fig:SU3_fit_RG} \footnote{In this section, when discussing the running coupling of the CF model for SU(2), we refer to the running coupling employed for the symmetric and orthogonal configurations, which differs from the one utilized in the soft gluon limit.}, denoted as $g_{\text{CF}}(\mu)$, with the running of the gauge coupling from \cref{eq:g_running}, $g_{\text{toy}}(\mu)$, with parameters $\Lambda$ and $g_0$, obtained from the fits for the gauge groups SU(2) and SU(3). We observe that $g_{\text{CF}}(\mu)$ and $g_{\text{toy}}(\mu)$ differ at all scales and, although the difference tends to diminish at high momenta, it is not negligible in the UV region.

An interesting analysis involves assessing how the differences among the UV tails of $G^{CCA}$ vary when we minimize the discrepancy between $g_{\text{CF}}(\mu)$ and $g_{\text{toy}}(\mu)$ in the UV domain. For this purpose, it is convenient to introduce the quantity $\chi_{g,UV}$, as a way to estimate that discrepancy:

\begin{equation}
    \chi_{g,UV}^{2}\equiv\sum_{p_i=6\ \text{GeV}}^{{p_i=10\ \text{GeV}}}\Bigg(\frac{g_{\text{toy}}(p_i)-g_{\text{CF}}(p_i)}{g_{\text{CF}}(p_i)}\Bigg)^2.
\end{equation}

\begin{figure}[h!]
    \centering
    \begin{subfigure}{0.4\textwidth}
    \includegraphics[width=\textwidth]{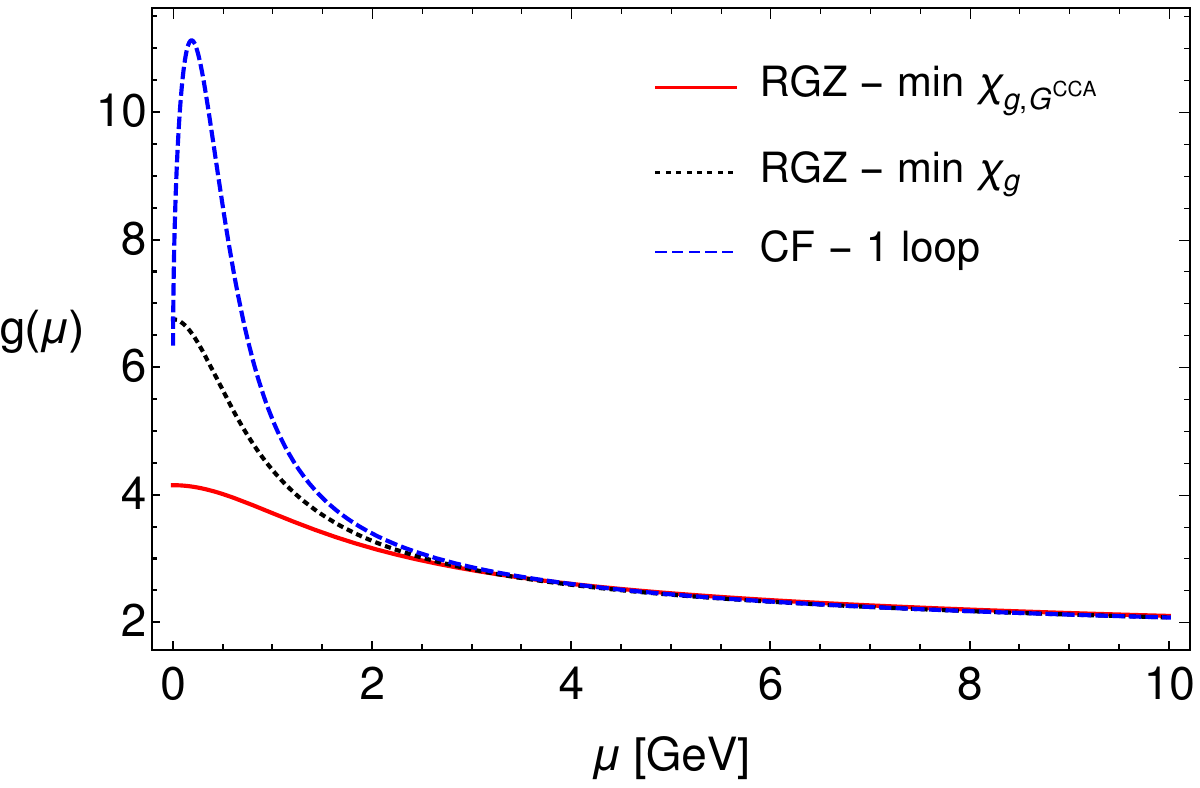}
    \end{subfigure}
    \begin{subfigure}{0.4\textwidth}
        \includegraphics[width=\textwidth]{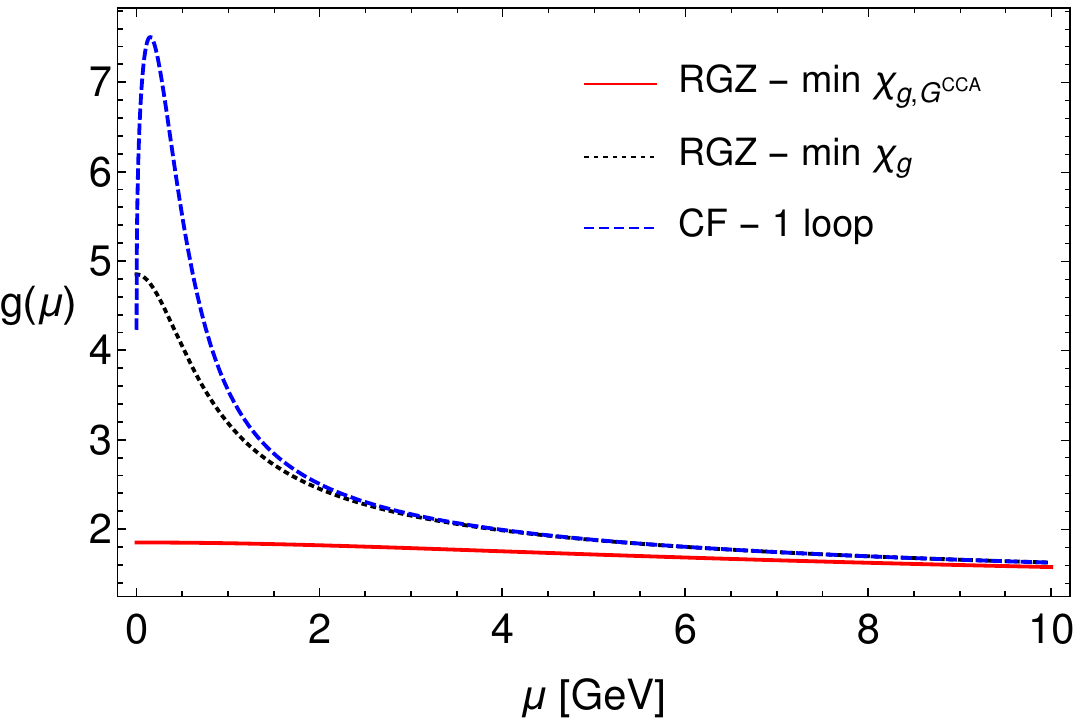}
    \end{subfigure}
    \caption{The running coupling $g_{\text{toy}}$ and $g_{\text{CF}}$ for the SU(2) (left) and SU(3) (right) gauge groups. The black curve corresponds to the parameters that minimize $\chi_{g,UV}^{2}$. These are $\Lambda=1.05$  GeV, $g_0=6.75$ and $\Lambda=0.90$  GeV, $g_0=4.85$, for SU(2) and SU(3) respectively. The red curve corresponds to the parameters that minimize $\chi_{g,G^{CCA}}$. These are $\Lambda=1.65$  GeV, $g_0=4.15$ and $\Lambda=5.05$  GeV, $g_0=1.85$, for SU(2) and SU(3) respectively.}
    \label{fig:g_best}
\end{figure}

To carry out the summation we used uniform steps of length $\Delta p=p_{n+1}-p_n=0.5$ GeV. The parameters that minimize $\chi_{g,UV}$ are $\Lambda=1.05$ GeV, $g_0=6.75$ for the SU(2) gauge group and $\Lambda=0.90$ GeV, $g_0=4.85$ for SU(3). These parameters lead to an excellent agreement between $g_{\text{toy}}(\mu)$ and  $g_{\text{CF}}(\mu)$ in the UV, as is illustrated in \cref{fig:g_best} (black curve). Unfortunately, these values do not feature good compatibility with the lattice data of $G^{CCA}$.  

To assess the potential for enhancing agreement between the results from the CF model and DSE with the RGZ framework without significantly compromising the IR description of lattice data, we introduce the following collective error

\begin{equation}
    \chi_{g,G^{CCA}}^2=\frac{1}{2}\Big(\chi_{g,UV}^{2}+\chi_{G^{CCA}}^2\Big),
\end{equation}
where $\chi_{G^{CCA}}$ is defined according to \cref{eq:joint_error} and \cref{eq:rel_error} for the SU(2) and SU(3) gauge groups, respectively. The quantity $\chi_{g,G^{CCA}}$ takes into account both the discrepancy between  $g_{\text{CF}}(\mu)$ and $g_{\text{toy}}(\mu)$ in the UV as well as the discrepancy between the dressing function $G^{CCA}$ from the RGZ theory and the corresponding lattice data.  

\begin{figure}[h!]
    \centering
    \begin{subfigure}{0.4\textwidth}
    \includegraphics[width=\textwidth]{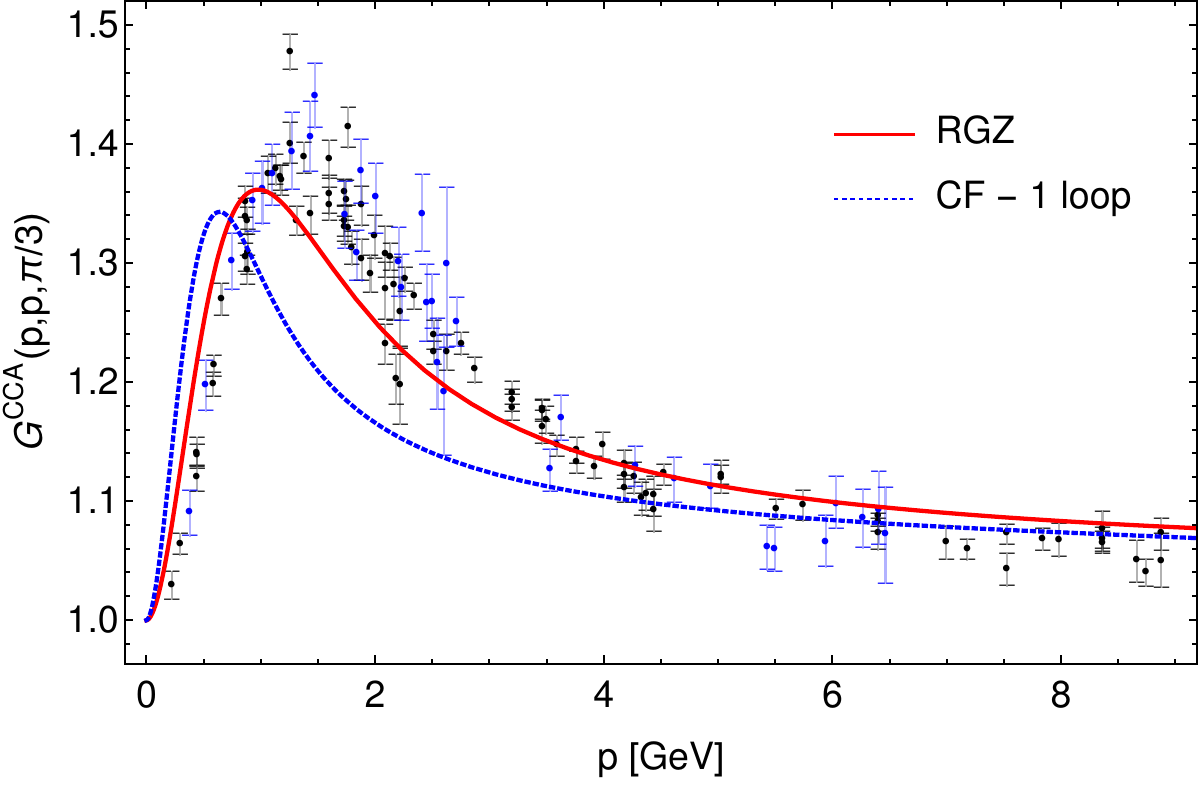}
    \end{subfigure}
    \begin{subfigure}{0.4\textwidth}
        \includegraphics[width=\textwidth]{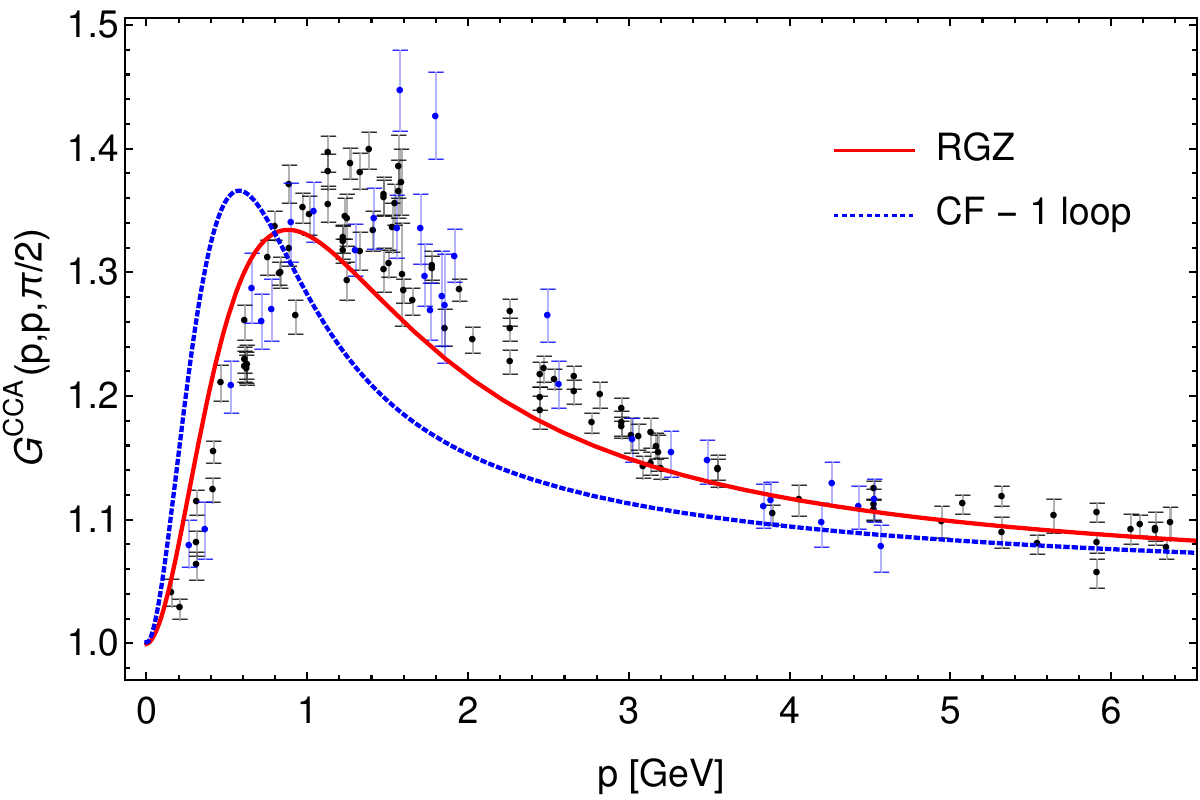}
    \end{subfigure} 
    \begin{subfigure}{0.4\textwidth}
        \includegraphics[width=\textwidth]{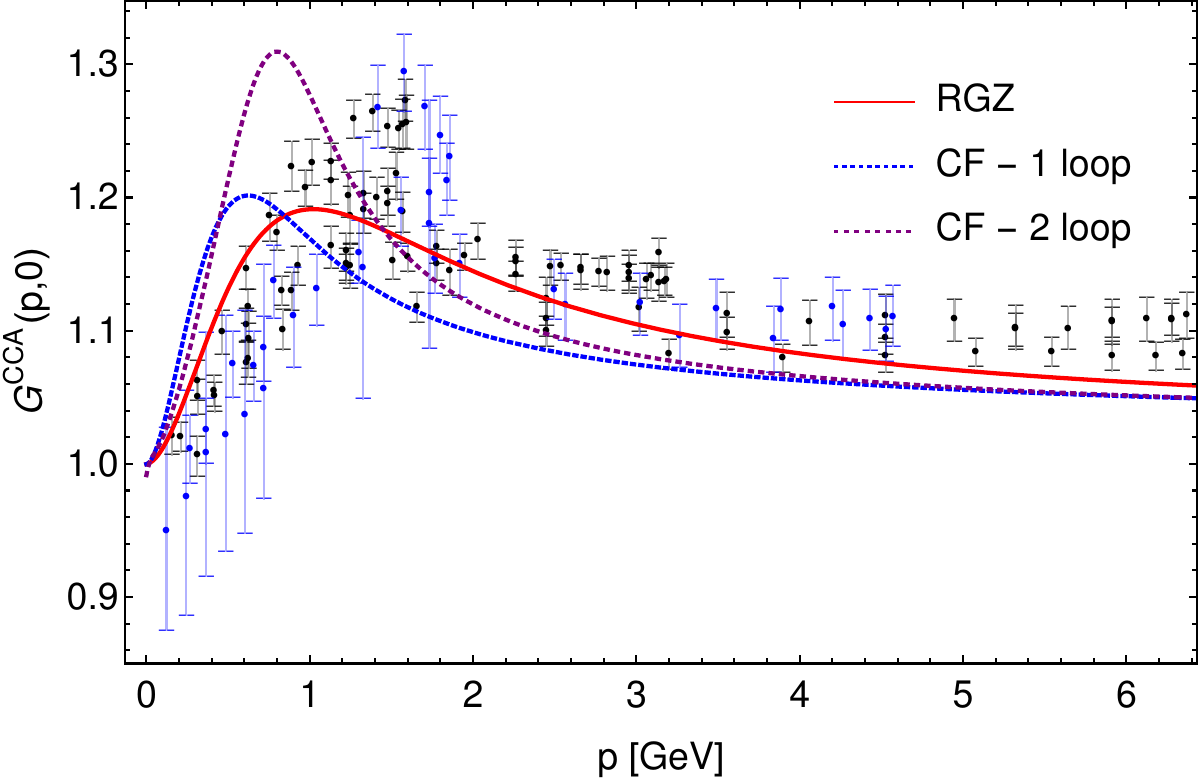}
    \end{subfigure} 
    \caption{The scalar function $G^{CCA}$ as a function of the antighost external momentum in three distinct configurations: the symmetric (top left), orthogonal (top right) and the soft gluon limit (bottom) in the case of the SU(2) theory. The gauge coupling is described by the toy model of \cref{eq:g_running} with parameters that minimize $\chi_{g,G^{CCA}}$, $\Lambda=1.65$  GeV and $g_0=4.15$.}
    \label{fig:SU2_chi_gCCA}
\end{figure}

The parameters $\Lambda$ and $g_0$, derived from minimizing $\chi_{g,G^{CCA}}$, would in principle generate a gauge coupling $g_{\text{toy}}(\mu)$ that balances both the UV alignment with $g_{\text{CF}}(\mu)$ and the consistency with lattice data across all momentum scales. Such running coupling is also depicted in \cref{fig:g_best} (red curve). In the case of the SU(2) gauge group, we notice an enhancement in the agreement in the UV region between RGZ and CF/DSE. Notably, this improvement does not imply a significant deterioration in the IR description. This is evident when comparing \cref{fig:SU2_chi_gCCA} with \cref{fig:SU2_fit_RG}.

In the case of SU(3), the minimization of $\chi_{G^{CCA}}$ leads to a running of the gauge coupling which is essentially constant. This is consistent with the plots of \cref{fig:SU2_fit_RG,fig:SU3_fit_RG}, where a fixed value of $g$ was capable of describing the lattice data for the entirety range of momentum. However, the aforementioned running does not reduce the disagreement between the results from the CF model and DSE with the RGZ framework outcomes in the UV, being the quality of the fits similar to the ones presented in \cref{fig:SU2_fit_RG,fig:SU3_fit_RG}. This is reasonable, since we are essentially neglecting the dependence of $g$ with the scale. Still, the results for SU(3) should be taken with care due to the dispersion of lattice data in the soft gluon limit and the lack of data in alternative kinematic setups.

We can conclude that the specific toy model we employed has an impact on the differences between the RGZ framework and other approaches in the UV region. However, these discrepancies probably have other sources as well such as the neglect of the RG flow of the remaining parameters of the theory and the specific choice of the renormalization scheme.

\newpage
\section{Summary and final remarks} 
\label{sec:finalremarks}
 
In this work we evaluated the one-loop ghost-gluon correlation function within the RGZ framework for an arbitrary kinematic configuration and investigated the role of the running coupling. We performed the calculation for the pure gauge theory in a four dimensional Euclidean spacetime. With the purpose of comparing with other approaches, we specifically focused on three distinct kinematic configurations: the symmetric and orthogonal configurations and the soft gluon limit. 

This calculation is important since it serves as a benchmark to assess how well the RGZ framework describes the behavior of YM theory in the low momentum regime. In this sense this paper comes to complement previous investigations on the two-point functions at tree-level order \cite{Dudal:2008sp,Dudal:2010tf} and the analysis of the ghost-gluon vertex in the case of the soft-gluon limit at one-loop order \cite{Mintz:2017qri}.

The only parameter of the theory we used to fit our results to the available lattice data for the ghost-gluon vertex was the gauge coupling. All the other parameters were extracted from refs. \cite{Cucchieri:2011ig} and \cite{Oliveira:2012eh}, where they were determined by fitting lattice results for the gluon propagator using the RGZ tree level expression. This choice reduces the number of free parameters to the minimum, making the present analysis a more strict test of the RGZ framework.

As already pointed out in \cite{Mintz:2017qri}, as for the soft gluon limit the results are compatible with lattice simulations for both SU(2) \cite{Cucchieri:2008qm,Maas:2019ggf} and SU(3) \cite{Ilgenfritz:2006he} by simply considering a constant value of the gauge coupling, \textit{i.e.} independent of the momentum scale. However, this is not the case of the symmetric and orthogonal configurations, where the renormalization group effects seem to be significant. Indeed, when introducing a toy model for the running of the gauge coupling, our outcomes are quantitatively compatible with available lattice data.   

Our results also display qualitative agreement with other continuum approaches, specifically the dynamical DSE solutions in two different truncation schemes \cite{Aguilar:2018csq} and the RG-improved CF model \cite{Pelaez:2013cpa,Barrios:2020ubx}. Additionally, we show that quantitative differences between the RGZ results and alternative continuum approaches in the UV region can be accounted for by the particularities of the model for the RG flow of the gauge coupling.  

Overall, these results support the RGZ theory as a valid description of the infrared YM dynamics. There are several ways, nevertheless, in which the present analysis can be improved. A fully dynamical calculation of the two-point functions of the RGZ framework, most likely in an infrared safe scheme, would allow us to determine the running parameters of the theory by fitting the two-point functions to lattice data. Another potential study concerning this correlation function could investigate its dependence on the gauge parameter within the context of linear covariant gauges, utilizing the BRST invariant version of the RGZ action \cite{Capri:2016aqq}.

\section*{Acknowledgements}

The authors acknowledge useful discussions with M. Tissier and N. Wschebor. The authors also would like to thank A. C. Aguilar and A. Maas for discussions and for kindly providing the data for the comparison plots in Section \ref{sec:discussion}.
This work has been partially supported by CNPq, CAPES, 
FAPERJ, PEDECIBA and the ANII-FCE-166479 project. It is a part of the project INCT-FNA Proc. 
464898/2014-5. M. S. Guimaraes is a CNPq researcher under contract 310049/2020-2. This study was financed in part by the Coordena\c{c}\~ao de Aperfei\c{c}oamento de Pessoal de N\'\i vel Superior - Brasil (CAPES) - Finance Code 001.

\appendix 

\section{Relevant Feynman rules (Landau gauge)}\label{sec:appendix-feynman-rules}

Given its many fields and interactions, the RGZ theory has a large number of propagators and vertices. 
However, for the calculation of the ghost-gluon vertex at one-loop level (and in the Landau gauge), only a few 
of them are required. The Feynman rules corresponding to these propagators and vertices are shown below.

\subsection{Tree-level propagators}

In order to calculate the ghost-gluon vertex function in the 
Refined Gribov-Zwanziger theory, only a subset of the propagators of 
the theory are needed. These are 
\begin{eqnarray}
\langle A^a_{\mu}(p)A^b_{\nu}(-p)\rangle &=& \delta^{ab}\left[\frac{p^2+M^2}{p^4+(m^2+M^2)p^2+m^2M^2+2Ng^2\gamma^4}
P^{\perp}_{\mu\nu}(p)\right]\equiv \delta^{ab}P^{\perp}_{\mu\nu}(p)D_{AA}(p)\label{eq:RGZgluonpropagator}\\\nonumber
\\
\langle A_{\mu}^a(p)\varphi_{\nu}^{bc}(-p)\rangle
&=&\frac{-ig\gamma^2f^{abc}}{p^4+p^2(m^2+M^2)+m^2M^2+2Ng^2\gamma^4}P^{\perp}_{\mu\nu}(p) 
= -ig\gamma^2f^{abc}P^{\perp}_{\mu\nu}(p)\frac{D_{AA}(p)}{p^2+M^2}
\\
\langle A_{\mu}^a(p)\bar{\varphi}_{\nu}^{bc}(-p)\rangle
&=&\frac{-ig\gamma^2f^{abc}}{p^4+p^2(m^2+M^2)+m^2M^2+2Ng^2\gamma^4}P^{\perp}_{\mu\nu}(p)=\langle A_{\mu}^a(p)\varphi_{\nu}^{bc}(-p)\rangle\\
\langle\bar{c}^a(p)c^b(-p)\rangle &=&\frac{1}{p^2}\delta^{ab}\equiv \delta^{ab}D_{\bar cc}(p)
\;
\end{eqnarray}
where
\begin{eqnarray}
 P^{\perp}_{\mu\nu}(p) = \delta_{\mu\nu} - \frac{p_\mu p_\nu}{p^2}
\end{eqnarray}
is the transverse projector, such that $p_\mu P^{\perp}_{\mu\nu}(p)=p_\nu P^{\perp}_{\mu\nu}(p)=0$. 

It is often convenient to write the $D_{AA}$ form factor as a sum of massive propagators, i.e.,
\begin{eqnarray}
D_{AA}(p^2) &=& \frac{p^2+M^2}{(p^2+m^2)(p^2+M^2)+\lambda^4} 
\equiv \frac{R_+}{p^2+a_+^2} + \frac{R_-}{p^2+a_-^2},
\end{eqnarray}
with
\begin{align}
 & a_+^2 = \frac{m^2+M^2+\sqrt{(m^2-M^2)^2-4\lambda^4}}{2},\nonumber\\
 & a_-^2 = \frac{m^2+M^2-\sqrt{(m^2-M^2)^2-4\lambda^4}}{2},\nonumber\\
 & R_+ = \frac{m^2-M^2+\sqrt{(m^2-M^2)^2-4\lambda^4}}{2\sqrt{(m^2-M^2)^2-4\lambda^4}},\nonumber\\
 & R_- = \frac{-m^2+M^2-\sqrt{(m^2-M^2)^2-4\lambda^4}}{2\sqrt{(m^2-M^2)^2-4\lambda^4}} = 1-R_+,
 \label{eq:poles_residues_AA}
\end{align}
with $\lambda^4 = 2Ng^2\gamma^4$.

Note that either the poles $a_+$ and $a_-$ are complex, or the residues $R_+$ and $R_-$ have opposite signs (so that one of them is negative). This property can be directly related to positivity violation of the gluon propagator in the RGZ theory.

\subsection{Tree-level vertices}

The only vertices needed for the computation of the ghost-gluon 
at one-loop in the RGZ theory are
\begin{eqnarray}
 ^{\rm tree}[\Gamma_{AAA}(k,p,q)]^{abc}_{\mu\nu\rho}&=&-\left.\frac{\delta^3S_{\rm tree}}{\delta A_\mu^a(k)\delta A_\nu^b(p)\delta A_\rho^c(q)}\right|_{\Phi=0} 
 = igf^{abc}\left[(k_\nu-q_\nu)\delta_{\rho\mu} + (p_\rho-k_\rho)\delta_{\mu\nu} + (q_\mu-p_\mu)\delta_{\nu\rho} \right] \,,\nonumber\\
^{\rm tree}[\Gamma_{A\bar c c}(k,p,q)]^{abc}_{\mu}&=& -\left.\frac{\delta^3S_{\rm tree}}{\delta A_\mu^a(k)\delta \bar c^{b}(p) \delta c^{c}(q) }\right|_{\Phi=0} = -igf^{abc}p_\mu\,,\nonumber\\
 ^{\rm tree}[\Gamma_{A\bar \varphi \varphi}(k,p,q)]^{abcde}_{\mu\nu\rho}&=& -\left.\frac{\delta^3S_{\rm tree}}{\delta A_\rho^a(k)\delta \bar\varphi_\mu^{bc}(p) \delta\varphi_\nu^{de}(q) }\right|_{\Phi=0} = -igf^{abd}\delta^{ce}\delta_{\nu\rho}p_\mu\,. \nonumber\\
\end{eqnarray}

Note that, for higher orders in the perturbative expansion, or for a general 
linear covariant gauge, or for other correlation functions, extra correlators 
and vertices will be needed.

\section{On the relation between connected and 1PI correlation functions in the presence of mixed propagators}\label{sec:appendix-mixed-propagators}

Let us denote the generating 
functional of connected correlation functions as $W[\vec J]$, where $J_i$ are external sources associated 
with the different elementary fields, and let $\Gamma[\vec\phi]$ be the quantum action, that is, the 
generating functional of 1PI correlation functions. Using this notation, we start from the well-known 
relation
 \begin{eqnarray}\label{eq:2point-connected-1PI-J}
\left.\frac{\delta^2\Gamma[\vec\phi]}{\delta\phi_j\delta\phi_\ell}\right|_{\vec \phi = \vec{\Phi}[\vec J]}\frac{\delta^2W[\vec J]}{\delta J_\ell\delta J_k} 
 = -\delta_{jk} 
\end{eqnarray}
Taking a further derivative with respect to the source $J_i$, one finds
\begin{eqnarray}\label{eq:connected-3-point}
 \frac{\delta^3W[\vec J]}{\delta J_i\delta J_p\delta J_k}&=& 
 -\frac{\delta^2W[\vec J]}{\delta J_p\delta J_j}\left(\left.\frac{\delta^3 
 \Gamma[\vec \phi]}{\delta \phi_j\delta \phi_\ell\delta\phi_m}\right|_{\vec \phi = \vec{\Phi}[\vec J]}\right)
 \frac{\delta^2W[\vec J]}{\delta J_i\delta J_m}\frac{\delta^2W[\vec J]}{\delta J_\ell\delta J_k}.
\end{eqnarray}

For the present calculation of the ghost-gluon vertex, we are specifically interested in the choice
\begin{eqnarray}
  i &=& A_\mu^e(k)\nonumber\\
  p &=& \bar c^a(p)\nonumber\\
  k &=& c^b(q).
\end{eqnarray}

Since there are no mixed propagators involving the Faddeev-Popov ghosts $c$ and $\bar c$, the only nonvanishing 
contributions are such that $j=c$ and $\ell=\bar c$. Therefore, running the remaining sum for 
$m=A,\varphi,\bar\varphi$, we have 
 \begin{eqnarray}
  \frac{\delta^3W[\vec J]}{\delta J_A\delta J_{\bar c}\delta J_c}&=& 
 -\frac{\delta^2W[\vec J]}{\delta J_c\delta J_{\bar c}}
 \left\{\left(\left.\frac{\delta^3 \Gamma[\vec \phi]}{\delta c\,\delta {\bar c}\,\delta A}\right|_{\vec \varphi = \vec{\Phi}[\vec J]}\right)
 \frac{\delta^2W[\vec J]}{\delta J_A\delta J_A} + \right.\nonumber\\
 &&+\left.\left(\left.\frac{\delta^3 \Gamma[\vec \phi]}{\delta c\,\delta \,{\bar c}\,\delta\varphi}\right|_{\vec \varphi = \vec{\Phi}[\vec J]}\right)
 \frac{\delta^2W[\vec J]}{\delta J_A\delta J_\varphi} + \right.
  \left.\left(\left.\frac{\delta^3 \Gamma[\vec \phi]}{\delta c\,\delta \,{\bar c}\,\delta\bar\varphi}\right|_{\vec \varphi = \vec{\Phi}[\vec J]}\right)
 \frac{\delta^2W[\vec J]}{\delta J_A\delta J_{\bar\varphi}}\right\}
 \frac{\delta^2W[\vec J]}{\delta J_{\bar c}\delta J_c},
 \end{eqnarray}
which, at the one-loop level, can be written as
\begin{eqnarray}
 \vev{A_\nu^a(k)\, \bar c^b(p)\,c^c(q)} = D_{\bar cc}(p)D_{\bar cc}(q) D_{AA}(k)P_{\mu\nu}^\perp(k)\left\{\frac{\delta^3\Gamma}{\delta A_\mu^a(-k)\delta \bar c^b(-p)\delta c^c(-q)}
 -
 \frac{2ig\gamma^2f^{ade}}{k^2+M^2}\frac{\delta^3\Gamma}{\delta c^b(-p)\delta \bar c^c(-q)\delta \varphi_\mu^{de}(-k)}\right\},\nonumber\\
\end{eqnarray}

where we used the fact that the correlation functions obey $\vev{A\varphi}=\vev{A\bar\varphi}$, and also $\delta^3\Gamma/\delta c\delta \bar c \delta\varphi=\delta^3\Gamma/\delta c\delta \bar c \delta\bar\varphi$.

In a shorthand notation, one can conveniently write
 \begin{equation}\label{eq:Acc-local}
  \frac{\vev{A\,\bar c\, c}_c}{(\vev{\bar c\,c}_c)^2\vev{AA}_c} = \Gamma_{A\,\bar c\,c} + \frac{\vev{A\,\varphi}_c}{\vev{A\,A}_c}\Gamma_{\bar c\,c\,\varphi} 
  +\frac{\vev{A\,\bar\varphi}_c}{\vev{A\,A}_c}\Gamma_{\bar c\,c\,\bar\varphi}.
 \end{equation}

Therefore, besides the contribution $\Gamma_{A\bar cc}$, that would be present at pure Yang-Mills, there are also contributions from 1PI functions involving the auxiliary fields $\varphi$ and $\bar\varphi$, as well as the respective mixed propagators. Such contributions can be thought as momentum-dependent contributions to the ghost-gluon vertex, which are present starting from the tree level of the RGZ action.

\section{Analytic expression for the ghost-gluon vertex in the soft-gluon limit}\label{sec:appendix-soft-gluon}

In the soft-gluon limit (i.e., when the gluon momentum $k\rightarrow0$), the expression (\ref{eq:Acbarc-tensor-decomp}) for the ghost-gluon vertex is relatively simpler than in a general kinematic regime. Given the absence of IR-divergences, the soft-gluon limit for the ghost-gluon vertex reads
\begin{eqnarray}
      \Gamma_{A^a_\mu\bar c^b c^c}(0,p,-p)=-igf^{abc}p_\mu B_1(0,p).
\end{eqnarray}

The scalar function $B_1(0,p)$ has been calculated (with a slightly different notation) to one-loop order in \cite{Mintz:2017qri}. The result can be written as\footnote{There is a typo in the corresponding result from \cite{Mintz:2017qri}, which is corrected in this expression.}
\begin{eqnarray}
    B_1(0,p) &=& 1 + \frac{Ng^2}{2}\left[R_+J(a_+;p) + R_-J(a_-;p)\right]
    - Ng^2\left[R_+^2K(a_+,a_+;p) + R_-^2K(a_-,a_-;p)+2R_+R_-K(a_+,a_-;p)\right]\nonumber\\
    &&+\frac{Ng^2}{2}\frac{Ng^2\gamma^4}{(a_+^2-a_-^2)^2}\left[K(a_+,a_+;p)+K(a_-,a_-;p)-2K(a_+,a_-;p)\right],
\end{eqnarray}
where the poles $a_\pm$ and residues $R_\pm$ of the tree-level gluon propagator are given by eq. (\ref{eq:poles_residues_AA}), and the scalar functions $J$ and $K$ above are given by
\begin{eqnarray}
J(m_1;p) 
  &=&\frac{1}{64\pi^2} \times\frac{2m_1^2p^2(p^2+m_1^2)+p^6\log\left(1+\frac{m_1^2}{p^2}\right)-(3p^2+2m_1^2)m_1^4\log\left(1+\frac{p^2}{m_1^2}\right)}{m_1^2p^4},    
\end{eqnarray}
\begin{eqnarray}
    K(m_1,m_2;p) &=& 
 \frac{1}{256\pi^2}\frac{1}{m_1^2m_2^2p^4(m_1^2-m_2^2)}\bigg\{2m_1^2m_2^6p^2-2m_1^6m_2^2p^2+3m_1^2m_2^4p^4 - 3m_1^4m_2^2p^4+\nonumber\\
&&\hspace{0.5cm}+\left.2m_1^8m_2^2\log\left(1+\frac{p^2}{m_1^2}\right)-2m_1^2m_2^8\log\left(1+\frac{p^2}{m_2^2}\right)+ 4m_1^6m_2^2p^2\log\left(1+\frac{p^2}{m_1^2}\right)-\right.\nonumber\\
&&\hspace{0.6cm}\left.- 4m_1^2m_2^6p^2\log\left(1+\frac{p^2}{m_2^2}\right) +
4m_1^2m_2^2p^6\log\left(\frac{p^2+m_2^2}{p^2+m_1^2}\right)+ \right.\nonumber\\
&&\hspace{0.7cm}\left.+
2m_1^2p^8\log\left(1+\frac{m_2^2}{p^2}\right) - 
2m_2^2p^8\log\left(1+\frac{m_1^2}{p^2}\right)\right\},
\end{eqnarray}
as for $m_1 \neq m_2$, and
\begin{equation}
    K(m,m;p)=\frac{-6 p^2 m^6-5 p^4 m^4-2p^6 m^2+\left(8 p^2 m^6+6m^8\right)\log \left(1+\frac{p^2}{m^2}\right)+2p^8 \log \left(1+\frac{m^2}{p^2}\right)}{256 \pi^2 p^4 m^4}.
\end{equation}

\newpage
\bibliography{RGZ-bibliography}

\end{document}